\documentclass[12pt,a4paper]{article}
\usepackage[T2A]{fontenc}
\usepackage[utf8]{inputenc}
\usepackage[english]{babel}
\usepackage{amssymb}
\usepackage{amsmath}
\usepackage{graphicx}
\usepackage{xcolor}
\usepackage{indentfirst}
\usepackage{siunitx}
\usepackage{longtable}

\def\affilmark#1{$^{#1}$}

\begin{document}

\begin{center}
	\textbf{Revision of calcium and scandium abundances in Am stars based on NLTE calculations
	        and comparison with diffusion stellar evolution models}

\renewcommand{\thefootnote}{\fnsymbol{footnote}}
\vskip 12pt
\textbf{L. I.~Mashonkina\affilmark{1}\footnote{E--mail: lima@inasan.ru}, Yu. A.~Fadeyev\affilmark{1}}

\vskip 12pt
\textit{$^1$Institute of Astronomy, Russian Academy of Sciences, Moscow, Russia}

\vskip 12pt
Received April 27, 2024; revised June 6, 2024; accepted June 6, 2024
\end{center}

\textbf{Abstract} ---
The homogeneous data sets for the calcium and scandium abundances accounting for departures from
local thermodynamic equilibrium were obtained for a sample of 54 metallic--line (Am) stars.
The  Са and Sc abundances were found to correlate with effective temperature $(T_\mathrm{eff})$,
the abundance growth with increasing $T_\mathrm{eff}$ being higher in stars with surface gravity
$\log g < 4$ than in those with $\log g\ge 4$.
No correlation was found between Ca or Sc abundances and the iron abundance or the velocity of axial
rotation.
Am stars exhibit on average the higher values of [Ca/H] than those of [Sc/H] as well as the abundance
ratio $[\textrm{Ca}/\textrm{Sc}] = 0.41\pm 0.30$.
However, at effective temperatures $T_\mathrm{eff} > 9500$\:K there is an allusion to the systematic
difference between Am stars with surface gravities $\log g \ge 4$ and $\log g < 4$.
The iron excess is nearly the same in the range $7200\:\textrm{K}\le T_\mathrm{eff}\le 10030\:\textrm{K}$.
Evolution diffusion models computed with the code \textsc{MESA} for stars with masses from 1.5 to $2M_\odot$
show the surface abundances that are in good agreement with Ca and Fe abundances observed in Am stars
of the three open clusters with the age greater than 600 Myr.
Additional mechanisms of chemical separation should be considered for explanation of the Am phenomenon
in young stars of the Pleiades cluster.
We tested the published diffusion stellar evolution models.
The diffusion models by Richer et al. (2000) and Hui--Bon--Hoa et al. (2022) are shown to agree with
observations of Am stars in the open clusters at large values of the free turbulence parameter:
$\omega\sim 10^3$ for Ca and Fe, $\omega\sim 500$ for Sc.
There is no model with the mass and age of the Am-type star Sirius that could reproduce its
surface abundances from He to Ni.
The results presented in the paper may be of importance for understanding the chemical peculiarity
of Am stars.


\setcounter{footnote}{0}

\newpage
\section{Introduction}

The masses of spectral type A stars range from 1.7 to $3M_\odot$, whereas their ages are from
$10^8$ to $2.7\times 10^9$ yr.
Therefore, as the most of Galactic thin disk stars they formed in the interstellar environments
with composition close to the solar one.
However observations exhibit strong inhomogeneity of A--type star elemental abundances
as well as the large deviations from the solar composition in some groups of A stars.
Subgroups of metallic--line (Am) stars, magnetic chemically--peculiar (Ap) stars,
$\lambda$ Boo--type stars were separated from superficially normal A--type stars on
the base of comparisons of their spectra.
Spectra of superficially normal A stars differ from the solar spectrum due to different values
of the effective temperature $T_\mathrm{eff}$ and surface gravity $\log g$ whereas their chemical
composition is close to the solar one.
Ap stars with values of $T_\mathrm{eff}$ and $\log g$ close to those of the normal A stars differ
from them due to significantly stronger lines of Si and Cr as well as of elements beyond the iron
group in the Mendeleev table.
All Ap stars possess strong magnetic fields ranging from 1 to 30 kG (Preston 1974).
Non--magnetic stars with strong lines of the iron--group elements and with weak Ca and Sc lines
are classified as Am stars (Conti 1970).
$\lambda$ Boo--type stars have the abundances of C, N and O nearly the same as solar abundances
but at the same time show deficiency of Mg, Al, Si, S, Mn, Fe and Ni (Baschek and Slettebak 1988).

Differences in chemical composition are mostly confined to the superficial layers of the star.
The peculiar composition of $\lambda$ Boo--type stars is interpreted as a result of accretion
of the interstellar gas with low abundance of elements participating in dust formation
(Venn and Lambert 1990).
For explanation of chemical peculiarity in Ap stars Michaud (1970) proposed the atomic diffusion
model of element separation due to gravitational settling and radiation pressure.
Stellar magnetic fields inhibit the mixing.
Watson (1970) expanded on this idea to explain the phenomenon of Am stars.
Similarly to Ap stars, Am stars are slow rotators with the rotational velocity as high as 120 km/s
(Jaschek and Jaschek 1957; Abt and Moyd 1973; Abt and Morrell 1995)
so that material mixing in their surface layers is not so efficient as in normal rapidly
rotating A stars with prominent meridional circulation.
Despite the long history of peculiar A star investigations the appropriate interpretation of
varieties of peculiar properties in their observed elemental abundances has not been achieved yet.
These are not only $\lambda$ Boo--type stars accreting the interstellar gas and Ap stars with
strong magnetic fields but also more ``simple'' Am stars.

It is the Am stars that are the goal of our study and where the Fe, Ca and Sc abundances are in the
focus of attention.
According to Conti (1970) the Fe, Ca and Sc abundances play a key role in classifying a star as Am.
It is these elements that seem to be responsible for chemical  peculiarity in Am stars.
We do not consider the elements beyond the iron group since their enhanced abundances compared to
the solar abundances are typical not only for Am stars but also for A stars though with fairly
lower degree (Mashonkina et al. 2020).
There are two sides of the Am star problem.
First, the accuracy and sufficiency of observational data
on elemental abundances in Am stars of different stellar parameters, such as
effective temperatures $T_\mathrm{eff}$, stellar masses, rotational velocities $V \sin i$, ages
and binarity.
Second, theoretical predictions of elemental abundances in the atmospheres of Am stars
based on stellar evolution calculations with effects of atomic diffusion taken into account.
Results of these calculations are called in the literature as diffusion models.

\begin{figure}
	\centering
	\includegraphics[width=0.45\columnwidth,clip]{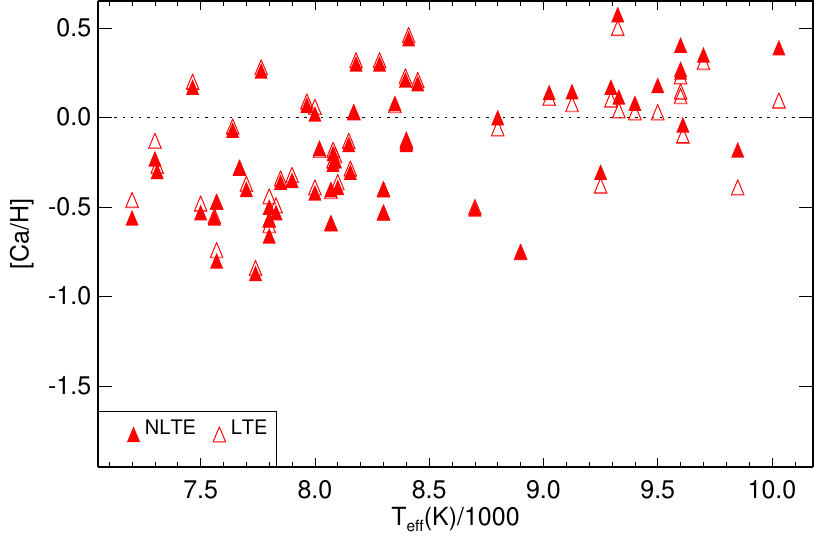}
    \includegraphics[width=0.45\columnwidth,clip]{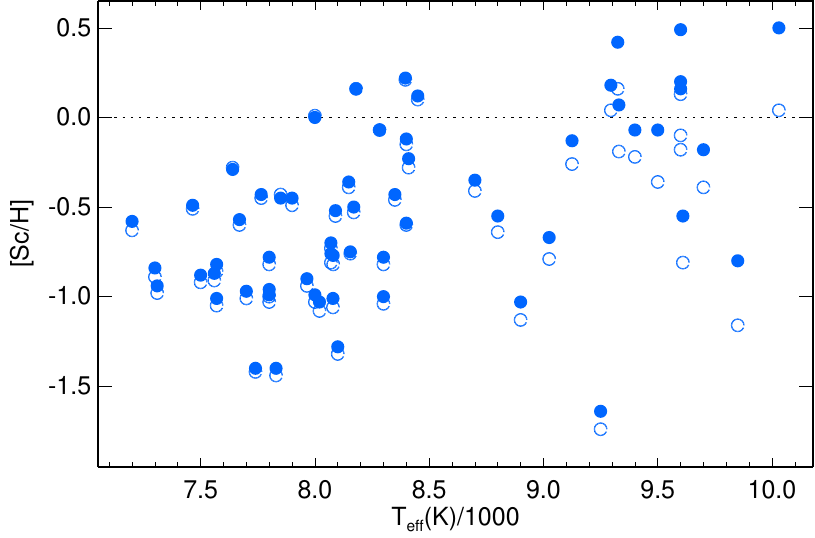}
	\caption{NLTE (filled symbols) and LTE (open symbols) abundances [Ca/H] and [Sc/H]
	in atmospheres of Am stars as a function of $T_\mathrm{eff}$.}
	\label{fig1}
\end{figure}

Renson and Manfroid (2009) published the catalogue involving 4299 stars that are identified or
suspected as Am and Fm stars according to the criteria of Conti (1970).
The new catalogue of chemically peculiar stars by Ghazaryan et al. (2018) represent the published
data on atmosphere parameters and abundances determined from the high resolution spectra for 129 stars.
The catalogue of Ghazaryan et al. (2018) was the base of our study but we did not consider Fm stars
because their outer convection zones may significantly affect chemical peculiarity compared to
Am stars.
The abundance data [X/H]\footnote{For any two elements X and Y:
$[\textrm{X}/\textrm{Y}] = \log (N_\mathrm{X}/N_\mathrm{Y})_\mathrm{star} -
\log (N_\mathrm{X}/N_\mathrm{Y})_\mathrm{Sun}$.}
catalogued by Ghazaryan et al. (2018) cannot be directly compared to the diffusion models since they
were obtained with different oscillator strengths $\log gf$ for the same spectral lines,
correspond to different estimates of solar abundances and are based on the assumption of local
thermodynamic equilibrium (LTE).
Sitnova et al. (2018) and Mashonkina (2024) showed that for lines Ca~I, Ca~II and Sc~II
departures from LTE (NLTE effects) strongly depend on the effective temperature and for
$7\times 10^3\:\textrm{K} < T_\mathrm{eff} < 10^4\:\textrm{K}$
NLTE abundance corrections range from small negative to significant positive values and are
as high as 0.40 dex (Ca~I 4226\AA) and 0.48 dex (Sc II 4246\AA).
Fig.~\ref{fig1} shows the change of [Ca/H] and [Sc/H] ratios when the LTE assumption is
replaced by NLTE methods.
It should be noted that NLTE methods lead to the different abundance dependences on $T_\mathrm{eff}$.
NLTE abundances have been determined so far only for five Am stars:
HD~48915 (Sirius) and HD~72660 (Ca, Fe: Mashonkina et al. 2020, Sc: Mashonkina 2024),
HD~180347 (Ca, Sc: Trust et al. 2023),
$\theta$~Vir and $o$~Peg (Ca, Fe: Romanovskaya et al. 2023).
Thus, the revision of published data on Ca and Sc abundances in Am stars is of importance.

Richer et al. (2000), Vick et al. (2010), Hui--Bon--Hoa et al. (2022) calculated the diffusion models
that can be used for comparison with observations.
Campilho et al. (2022) computed the diffusion models of main sequence stars with masses $1M_\odot$ and
$1.4M_\odot$ using three different stellar evolution codes.
These are the Montreal/Montpellier evolution code (Turcotte et al. 1998),
the \textsc{CESAM} code (Morel and Lebreton 2008) and the \textsc{MESA} code (Paxton et al. 2018).
The small difference between the abundance profiles computed with use of these codes allows us
to conclude that all these computer programs are equally applicable for calculation of stellar evolution
with atomic diffusion.

In the present study we pursue the following goals.
\begin{itemize}
 \item Collection of homogeneous and accurate data on Ca, Sc and Fe abundances in the fairly
       extensive list of Am stars appropriate for tests of models of chemical peculiarity.
       In the present study we deal with published data,
       however, it was revised by accounting for NLTE effects.
       Abundances determined for individual spectral lines are reduced to the unified system
       of $\log gf$ values and the abundance [X/H] is evaluated using compositions of chondritic
       meteorite groups (Lodders 2021).
 \item The search for possible correlations between the chemical and physical stellar parameters:
       [Ca/H], [Sc/H], [Fe/H], $T_\mathrm{eff}$, $\log g$, $V\sin i$ and the age.
 \item Calculation of diffusion models for Ca and Fe abundances using the \textsc{MESA} code.
       Analysis of mechanisms responsible for chemical peculiarity of Am stars based on
       the comparison of diffusion models with observational data.
\end{itemize}

The paper is organized as follows.
In Section~2 we compile the sample of Am stars obtained from the published data.
Then, we determine the homogeneous LTE abundances and calculate Ca and Sc NLTE abundances
for stars of the sample.
Analysis of these data is presented in Section~3.
In Section~4 we compare observational abundances with published diffusion models as well as with
those we computed using the code \textsc{MESA}.
Finally we formulate our conclusions.

\section{NLTE abundances of Ca and Sc in Am stars}

\subsection{The sample of stars}

The catalogue by Ghazaryan at al. (2018) contains not only the parameters of stellar atmospheres
and abundances of chemical elements but also the references for data sources.
Among 129 stars mentioned in the catalogue as AmFm or uncertain we excluded the data for 55 stars.
These are the stars with undefined abundance of Sc, one star with undefined abundance of Ca,
one Fm star, one star with uncertain Am--type due to its high rotation velocity $V\sin i = 102$ km/s,
Vega which belongs to $\lambda$~Boo type--stars and three $\delta$~Sct stars where stellar pulsations
might affect the mechanisms of chemical peculiarity in comparison with ordinary Am stars.
Moreover, we excluded three stars with Ca and Sc abundances determined in original works on
the base of the differential approach for each spectral line relative to the corresponding
line of the solar spectrum due to the impossibility to evaluate the Ca and Sc absolute abundances.
We excluded three spectroscopic binary stars since their spectra do not allow us to
determine contribution of each component.
Finally, we excluded six stars due to large errors in the Ca and Sc abundance estimates
ranging from 0.3 to 0.57 dex.
Thus, the remaining list contains data for 55 Am stars.

The data were extended by six stars that are not present in the catalogue by Ghazaryan at al. (2018).
These are HD~94334 (Caliskan and Adelman 1997), HD~154029 (Adelman 1999),
HD~187959 and HD~202431 (Catanzaro et al. 2022),
HD~114330 (Romanovskaya et al. 2023) and HD~180347 (Trust et al. 2023).
Thus, our sample consists of 61 stars that are listed in Table~\ref{table1}.

{\setlength{\tabcolsep}{3pt}
\begin{longtable}{rrcrccrccrrrr}
\caption{Am stars: LTE and NLTE abundances of Ca and Sc}\label{table1}\\
\hline
HD & $T_\mathrm{eff}$ & $\log g$ & [Fe/ & \multicolumn{2}{c}{$\varepsilon{}$(Ca)} & [Ca/ & \multicolumn{2}{c}{$\varepsilon{}$(Sc)} & [Sc/ & [Ca/ & $V\sin i$ & Ref \\
\cline{5-6}
\cline{8-9}
   &  (K)  &      &  H]  & LTE & NLTE & H] & LTE & NLTE & H] & Sc] & (km/s) & \\
		\noalign{\smallskip} \hline \noalign{\smallskip}
\endfirsthead

\multicolumn{13}{c}{\tablename\ \thetable{} -- continued} \\
\hline
HD & $T_\mathrm{eff}$ & $\log g$ & [Fe/ & \multicolumn{2}{c}{$\varepsilon{}$(Ca)} & [Ca/ & \multicolumn{2}{c}{$\varepsilon{}$(Sc)} & [Sc/ & [Ca/ & $V\sin i$ & Ref \\
\cline{5-6}
\cline{8-9}
   &  (K)  &      &  H]  & LTE & NLTE & H] & LTE & NLTE & H] & Sc] & (km/s) & \\
		\noalign{\smallskip} \hline \noalign{\smallskip}
\endhead

\hline\endfoot

\hline\endlastfoot

22128A &  7560 & 4.00 & 0.48 & 5.72 & 5.71\scriptsize{(0.19)} & --0.56 & 2.13 & 2.17\scriptsize{(0.29)} & --0.87 &  0.31 & 19 &  1   \\
 48915 &  9850 & 4.30 & 0.53 & 5.88 & 6.09\scriptsize{(0.05)} & --0.18 & 1.88 & 2.24\scriptsize{(0.03)} & --0.80 &  0.62 & 16 &  2   \\
56495A &  7800 & 4.00 & 0.43 & 5.70 & 5.70\scriptsize{(0.29)} & --0.57 & 2.01 & 2.05\scriptsize{(0.40)} & --0.99 &  0.42 & 36 &  1   \\
 72660 &  9700 & 4.10 & 0.65 & 6.58 & 6.62\scriptsize{(0.09)} &   0.35 & 2.65 & 2.86\scriptsize{(0.04)} & --0.18 &  0.53 &  6 &  2   \\
89822B &  8900 & 4.20 & 0.04 & 5.52 & 5.52\scriptsize{(0.21)} & --0.75 & 1.91 & 2.01\scriptsize{(0.18)} & --1.03 &  0.28 & 18 &  3   \\
94334  & 10030 & 3.88 & 0.26 & 6.36 & 6.66\scriptsize{(0.14)} &   0.39 & 3.08 & 3.54\scriptsize{(0.07)} &  0.50 & --0.11 & 47 &  4   \\
95418  &  9600 & 3.80 & 0.29 & 6.39 & 6.54\scriptsize{(0.10)} &   0.27 & 2.86 & 3.20\scriptsize{(0.00)} &   0.16 &  0.11 & 46 &  5   \\
95608  &  9250 & 4.25 & 0.48 & 5.89 & 5.96\scriptsize{(0.14)} & --0.31 & 1.30 & 1.40$^1$                & --1.64 &  1.33 & 21 &  6,7 \\
114330 &  9600 & 3.61 & 0.38 & 6.50 & 6.68\scriptsize{(0.10)} &   0.41 & 3.17 & 3.53\scriptsize{(0.05)} &  0.49 & --0.09 & 18 &  8   \\
154029 &  9325 & 3.65 & 0.44 & 6.77 & 6.84\scriptsize{(0.16)} &   0.57 & 3.20 & 3.46\scriptsize{(0.10)} &   0.42 &  0.16 & 29 &  9   \\
173648 &  8155 & 3.90 & 0.38 & 5.98 & 5.97\scriptsize{(0.13)} & --0.31 & 2.28 & 2.29\scriptsize{(0.01)} & --0.75 &  0.45 & 47 &  6   \\
176843 &  7200 & 3.60 & 0.21 & 5.81 & 5.71\scriptsize{(0.17)} & --0.56 & 2.41 & 2.46\scriptsize{(0.08)} & --0.58 &  0.02 & 28 &  13  \\
180239 &  8100 & 4.00 & 0.36 & 5.91 & 5.88\scriptsize{(0.18)} & --0.39 & 1.72 & 1.76\scriptsize{(0.13)} & --1.28 &  0.89 & 42 &  13  \\
180347 &  7740 & 3.98 & 0.22 & 5.43 & 5.40\scriptsize{(0.24)} & --0.87 & 1.62 & 1.64\scriptsize{(0.11)} & --1.40 &  0.53 & 14 &  10  \\
182564 &  9125 & 3.80 & 0.41 & 6.35 & 6.41\scriptsize{(0.16)} &   0.15 & 2.78 & 2.91\scriptsize{(0.09)} & --0.13 &  0.28 & 25 &  11  \\
187254 &  8400 & 3.80 & 0.58 & 6.13 & 6.15\scriptsize{(0.08)} & --0.12 & 2.89 & 2.92\scriptsize{(0.05)} & --0.12 & --0.00 & 14 & 13  \\
187959 &  8000 & 4.50 & 0.53 & 6.33 & 6.29\scriptsize{(0.16)} &   0.02 & 3.05 & 3.04\scriptsize{(0.13)} & --0.00 &  0.02 & 43 &  14  \\
188911 &  7800 & 3.90 & 0.19 & 5.67 & 5.61\scriptsize{(0.25)} & --0.66 & 2.04 & 2.08\scriptsize{(0.14)} & --0.96 &  0.30 & 10 &  13  \\
190165 &  7300 & 3.80 & 0.50 & 6.14 & 6.04\scriptsize{(0.19)} & --0.23 & 2.15 & 2.20\scriptsize{(0.35)} & --0.84 &  0.61 & 61 &  13  \\
202431 &  7500 & 4.00 & 0.65 & 5.79 & 5.74\scriptsize{(0.22)} & --0.53 & 2.12 & 2.16\scriptsize{(0.16)} & --0.88 &  0.35 &  9 &  14  \\
209625 &  7700 & 3.65 & 0.16 & 5.90 & 5.87\scriptsize{(0.25)} & --0.40 & 2.03 & 2.07\scriptsize{(0.20)} & --0.97 &  0.57 & 36 &  12  \\
214994 &  9600 & 3.81 & 0.34 & 6.41 & 6.53\scriptsize{(0.14)} &   0.25 & 2.94 & 3.24\scriptsize{(0.13)} &   0.20 &  0.05 & 14 &  8   \\
225365 &  8800 & 3.80 & 0.26 & 6.21 & 6.27\scriptsize{(0.12)} &   0.00 & 2.40 & 2.49\scriptsize{(0.15)} & --0.55 &  0.55 & 40 &  13  \\
225410 &  7900 & 3.70 & 0.41 & 5.95 & 5.92\scriptsize{(0.04)} & --0.35 & 2.55 & 2.59\scriptsize{(0.10)} & --0.45 &  0.10 & 27 &  13  \\
225463 &  8300 & 3.80 & 0.12 & 5.87 & 5.87\scriptsize{(0.16)} & --0.40 & 2.00 & 2.04\scriptsize{(0.09)} & --1.00 &  0.60 & 13 &  13  \\
3134 1603$^2$ & 8300 & 3.90 & 0.30 & 5.74 & 5.74\scriptsize{(0.16)} & --0.53 & 2.22 & 2.26\scriptsize{(0.32)} & --0.78 & 0.25 & 36 & 13  \\
3143 1192$^2$ & 8000 & 3.80 & 0.15 & 5.88 & 5.85\scriptsize{(0.18)} & --0.42 & 2.01 & 2.05\scriptsize{(0.20)} & --0.99 & 0.57 & 15 & 13  \\
3143 1942$^2$ & 7800 & 3.90 & 0.06 & 5.83 & 5.77\scriptsize{(0.29)} & --0.50 & 2.22 & 2.26\scriptsize{(0.17)} & --0.78 & 0.28 &  8 & 13  \\
St1612 &  8400 & 4.10 & 0.44 & 6.14 & 6.12\scriptsize{(0.05)} & --0.15 & 2.44 & 2.45\scriptsize{(0.00)} & --0.59 &  0.44 & 68 &  15  \\
\multicolumn{13}{l}{The cluster NGC 6405}                                                                     \\
--32 13109$^3$ & 9400 & 4.20 & 0.43 & 6.30 & 6.35\scriptsize{(0.09)} & 0.08 & 2.82 & 2.97\scriptsize{(0.08)} & --0.07 & 0.15 &  5 &  16  \\
318091 &  8700 & 4.00 & 0.08 & 5.77 & 5.76\scriptsize{(0.09)} & --0.51 & 2.63 & 2.69\scriptsize{(0.08)} & --0.35 & --0.16 & 62 & 16  \\
\multicolumn{13}{l}{The cluster Coma$^4$, [Fe/H] = --0.01, the age 710 Myr}                              \\
107168 &  8283 & 4.20 & 0.44 & 6.59 & 6.57\scriptsize{(0.16)} &  0.30 & 2.97 & 2.97\scriptsize{(0.10)} & --0.07 &  0.37 & 14 &   17  \\
108486 &  8148 & 4.11 & 0.25 & 6.14 & 6.12\scriptsize{(0.15)} & --0.15 & 2.65 & 2.68\scriptsize{(0.31)} & --0.36 &  0.21 & 37 &  17  \\
108642 &  8079 & 4.06 & 0.21 & 6.03 & 6.01\scriptsize{(0.18)} & --0.26 & 1.98 & 2.03\scriptsize{(0.20)} & --1.01 &  0.75 &  9 &  17  \\
108651 &  8090 & 4.24 & 0.71 & 6.06 & 6.03\scriptsize{(0.15)} & --0.24 & 2.49 & 2.52\scriptsize{(0.00)} & --0.52 &  0.28 & 21 &  18  \\
\multicolumn{13}{l}{The cluster Pleiades$^4$, [Fe/H] = 0.00, the age 110 Myr}                                                           \\
22615 &  8410 & 3.83 & 0.20 & 6.73 & 6.71\scriptsize{(0.10)} &  0.44 & 2.76 & 2.81\scriptsize{(0.10)} & --0.23 &  0.67 & 30 &   19  \\
23325 &  7640 & 4.23 & 0.39 & 6.22 & 6.20\scriptsize{(0.30)} & --0.07 & 2.76 & 2.75\scriptsize{(0.22)} & --0.29 &  0.22 & 70 &  19  \\
23631 &  9610 & 4.34 & 0.31 & 6.17 & 6.23\scriptsize{(0.05)} & --0.04 & 2.23 & 2.49\scriptsize{(0.05)} & --0.55 &  0.51 & 10 &  19  \\
23924  &  8180 & 4.30 & 0.37 & 6.59 & 6.57\scriptsize{(0.15)} &  0.30 & 3.20 & 3.20\scriptsize{(0.00)} &  0.16 &  0.14 & 33 &    20  \\
\multicolumn{13}{l}{The cluster Hyades$^4$, [Fe/H] = 0.12, the age 800 Myr}                                                         \\
27628 &  7310 & 4.12 & 0.12 & 6.00 & 5.97\scriptsize{(0.00)} & --0.30 & 2.06 & 2.10\scriptsize{(0.00)} & --0.94 &  0.64 & 31 &  21  \\
27749  &  7570 & 4.30 & 0.61 & 5.53 & 5.47\scriptsize{(0.21)} & --0.80 & 1.99 & 2.03\scriptsize{(0.00)} & --1.01 &  0.21 & 16 &  20  \\
27962 &  9025 & 3.95 & 0.37 & 6.38 & 6.41\scriptsize{(0.00)} &   0.14 & 2.25 & 2.37\scriptsize{(0.00)} & --0.67 &  0.81 & 11 &  21  \\
28226 &  7465 & 4.09 & 0.36 & 6.47 & 6.44\scriptsize{(0.00)} &   0.17 & 2.53 & 2.55\scriptsize{(0.00)} & --0.49 &  0.66 & 83 &  21  \\
28355 &  7965 & 3.97 & 0.40 & 6.36 & 6.34\scriptsize{(0.00)} &   0.07 & 2.10 & 2.14\scriptsize{(0.00)} & --0.90 &  0.97 & 90 &  21  \\
28546 &  7765 & 4.20 & 0.16 & 6.55 & 6.53\scriptsize{(0.00)} &   0.26 & 2.59 & 2.61\scriptsize{(0.00)} & --0.43 &  0.69 & 28 &  21  \\
30210 &  8080 & 3.92 & 0.56 & 6.09 & 6.07\scriptsize{(0.00)} & --0.20 & 2.22 & 2.27\scriptsize{(0.00)} & --0.77 &  0.57 & 57 &  21  \\
33204  &  7670 & 4.00 & 0.24 & 5.99 & 5.99\scriptsize{(0.10)} & --0.28 & 2.44 & 2.47\scriptsize{(0.00)} & --0.57 &  0.29 & 36 &  22  \\
33254  &  7830 & 4.13 & 0.50 & 5.78 & 5.74\scriptsize{(0.16)} & --0.53 & 1.60 & 1.64\scriptsize{(0.00)} & --1.40 &  0.87 & 13 &  20  \\
\multicolumn{13}{l}{The cluster Praesepe$^4$, [Fe/H] = 0.16, the age 630 Myr}                                                                \\
73045  &  7570 & 4.05 & 0.56 & 5.80 & 5.80\scriptsize{(0.27)} & --0.47 & 2.18 & 2.22\scriptsize{(0.10)} & --0.82 &  0.35 & 10 &  18  \\
73174  &  8350 & 4.15 & 0.71 & 6.34 & 6.35\scriptsize{(0.08)} &   0.08 & 2.58 & 2.61\scriptsize{(0.08)} & --0.43 &  0.51 &  5 &  23  \\
73618  &  8170 & 4.00 & 0.46 & 6.30 & 6.30\scriptsize{(0.07)} &   0.03 & 2.51 & 2.54\scriptsize{(0.27)} & --0.50 &  0.53 & 47 &  23  \\
73709  &  8070 & 3.78 & 0.54 & 5.86 & 5.87\scriptsize{(0.06)} & --0.40 & 2.23 & 2.28\scriptsize{(0.01)} & --0.76 &  0.36 & 10 &  23  \\
73711  &  8020 & 3.69 & 0.15 & 6.09 & 6.10\scriptsize{(0.08)} & --0.17 & 1.96 & 2.01\scriptsize{(0.01)} & --1.03 &  0.86 & 62 &  23  \\
73730  &  8070 & 3.97 & 0.45 & 5.68 & 5.68\scriptsize{(0.03)} & --0.59 & 2.31 & 2.34\scriptsize{(0.03)} & --0.70 &  0.11 & 29 &  23  \\
\end{longtable}
\noindent
\textbf{Remarks.}
Numbers in parentheses are the r.m.s. errors $\sigma$.
 References to the sources of atmosphere parameters and observational data:
1 -- Folsom et al. (2013),
2 -- Mashonkina et al. (2020),
3 -- Adelman (1994),
4 -- Caliskan and Adelman (1997),
5 -- Adelman et al. (2011),
6 -- Adelman et al.  (1999),
7 -- this paper,
8 -- Romanovskaya et al. (2023),
9 -- Adelman (1999),
10 -- Trust et al. (2023),
11 -- Adelman (1996),
12 -- Adelman et al. (1997),
13 -- Niemczura et al. (2015),
14 -- Catanzaro et al. (2022),
15 -- Netopil et al. (2014),
16 -- K{\i}l{\i}{\c{c}}o{\u{g}}lu et al. (2016),
17 -- Gebran et al. (2008),
18 -- Hui-Bon-Hoa et al.  (1997),
19 -- Gebran and Monier (2008)
20 -- Hui-Bon-Hoa and Alecian (1998),
21 -- Gebran et al. (2010),
22 -- Varenne and Monier (1999),
23 -- Fossati et al. (2007).

\noindent
$^1$ the upper limit,

\noindent
$^2$ TYC,

\noindent
$^3$ CD,

\noindent
$^4$ the metallicity and age of the cluster according to Netopil et al. (2022).

\subsection{Stellar physical parameters}

For each star we use the values of $T_\mathrm{eff}$, $\log g$ and [Fe/H] presented in the papers
referenced in Table~\ref{table1}.
In several cases we compared the estimates of $T_\mathrm{eff}$ and $\log g$ obtained in the different
works and were convinced that they all agree within their errors.
For example, Adelman et al. (1999) have obtained $T_\mathrm{eff}=9250$\:K and $\log g=4.25$ for
HD~95608 whereas the later estimates by Khalack and LeBlanc (2015) are $T_\mathrm{eff}=9200$\:K and
$\log g=4.26$.
Similarly, for HD~214994 Romanovskaya et al. (2023) obtained $T_\mathrm{eff}=9600$\:K and
$\log g=3.81$ but estimates by Adelman et al. (2015) are $T_\mathrm{eff}=9535$\:K and $\log g=3.73$.
The values of [Fe/H] we recalculated with $\log\varepsilon_\mathrm{met,Fe}=7.45$ (Lodders 2021).
Hereafter we use the abundance scale with $\log\varepsilon(\textrm{H})=12$.
Values of $V\sin i$ we take from the catalogue of AmFm stars (Ghazaryan et al. 2018) and
in the case of missing data we used the data from the catalogue by Royer et al. (2007) or
from the papers referenced in Table~\ref{table1}.

In Section 4 we compare the diffusion models with Am stars observed in four star clusters.
Their ages and metallicities were taken from Netopil et al. (2022).

\subsection{Determination of the calcium and scandium abundances}

In order to obtain the homogeneous data for Ca and Sc abundances in Am stars we had to reduce
the data from various works to the common oscillatory strengths and to the common values
of the Ca and Sc solar abundances: $\log\varepsilon_\mathrm{met,Ca}=6.27$ and
$\log\varepsilon_\mathrm{met,Sc}=3.04$ (Lodders 2021).
The values of $\log gf$ were taken from VALD (Vienna Atomic Line Database,
Ryabchikova et al. 2015; Pakhomov et al. 2019).
The atomic energy levels of scandium are characterized by hyperfine splitting (HFS).
For lines of Sc~II VALD provides the data on hyperfine structure calculated with values of
$\log gf$ determined through laboratory measurements (Lawler et al. 2019).

The NLTE Ca abundance in five stars of our sample have been presented earlier
(Mashonkina et al. 2020; Romanovskaya et al. 2023; Trust et al. 2023).
Three of these stars have also estimates of the NLTE Sc abundance (Trust et al. 2023; Mashonkina 2024).
All these works were done with atomic parameters from VALD.

For remaining stars of our sample the NLTE abundances of Sc were evaluated by summation of the
LTE abundance and the NLTE correction
$\Delta_\mathrm{NLTE} = \log\varepsilon_\mathrm{NLTE} - \log\varepsilon_\mathrm{LTE}$.
For Ca~I and Ca~II lines we used the corrections calculated by Sitnova et al. (2018) spanning
over a wide range of stellar parameters whereas for Sc~II lines we used the corrections
computed by Mashonkina (2024).
The correction $\Delta_\mathrm{NLTE}$ corresponding to the individual line in the spectrum of
the star was evaluated by interpolation of correction grid values with respect to
$T_\mathrm{eff}$, $\log g$ as well as with respect to [Sc/H] for the Sc lines.

Adelman and his coauthors (eight papers cited below as Adelman+), Varenne and Monier (1999) and
K{\i}l{\i}{\c{c}}o{\u{g}}lu et al. (2016) presented both the spectral line lists and corresponding
$gf$ values.
Hui-Bon-Hoa et al. (two papers cited below as Hui-Bon-Hoa+) present only the spectral line lists
but we assumed that they used the same values of $gf$ as other authors cited above since all these
works were published almost at the same time.
LTE abundances determined in these studies for 18 stars were reduced to the actual $gf$ values of VALD.
The hyperfine structure of the Sc~II lines was taken into account only in the work of
K{\i}l{\i}{\c{c}}o{\u{g}}lu et al. (2016).
Neglect of HFS leads to the weaker 
computed Sc~II lines and thereby to the higher derived abundances.
Thus, neglect of HFS 
in the most of Sc abundance studies
cannot be responsible for its deficiency in Am stars.
In contrary, the Sc abundance in investigated stars seems to be slightly overestimated by
0.04, 0.03, 0.02, and 0.01~dex for $T_\mathrm{eff}=7000$, 8000, 9000 and $10^4$\:K in the case of
the strongest line Sc~II 4246\AA. 
For Sc~II 5526\AA, the error is 0.01~dex for $T_\mathrm{eff}=7000$\:K and becomes less
with increasing effective temperature.
It should be noted that these values correspond to the solar abundance of Sc.
Abundance of Sc in Am stars is lower than that of the Sun, the HFS plays a smaller role and we did not
correct the data for this effect.

In their studies Gebran and Monier (2008), Gebran et al. (2008, 2010) (below we cite as Gebran+)
and Royer et al. (2014) used the same list of lines.
We assume that the value $\log gf=-0.135$ for the line Ca~II 3933\AA\ (see Table~8 in Genran et al. 2008)
is erroneous due to misprint and should be replaced by $\log gf=0.135$.
And if so, the $\log gf$ values of Ca~II lines used by Gebran+ and in our study differ less than
0.01 dex.
For Sc~II lines the values of $\log gf$ in Gebran+ are higher than those in our study by 0.01
to 0.03 dex.
Gebran+ presented the mean LTE abundances of Ca and Sc for stars they investigated.
For Ca we do not introduce a correction
but for scandium its LTE abundance is reduced  by 0.02 dex for all 14 stars.
The NLTE corrections were calculated for individual Ca~II and Sc~II lines.
In each star the NLTE corrections corresponding to different Ca~II lines differ less than 0.02 dex.
The same conclusion is valid for lines of Sc~II with exception of Sc~II 4246\AA,
which has the greater $\Delta_\mathrm{NLTE}$ compared to the other lines, by 0.04 to 0.07~dex for different stars.
The largest disagreement between NLTE corrections for Sc~II 4246\AA\ and for other lines of Sc~II
is obtained when the theoretical equivalent width (EW) of Sc~II 4246\AA\ exceeds 120 m\AA.
To determine NLTE Ca and Sc abundances we used the mean values of $\Delta_\mathrm{NLTE}$.
It should be noted that for EW > 120 m\AA\ the line Sc~II 4246\AA\ was excluded.

Fossati et al. (2007) and Netopil et al. (2014) did not give the line lists but remarked that
they used the atomic parameters from VALD.
This implies that as in works by Gebran+ the values of $\log gf$ for Sc~II lines were taken from
Lawler and Dakin (1989).
Ca abundance does not need to be corrected
whereas Sc abundance was reduced by 0.02~dex for each of the seven stars.
Niemczura et al. (2015) and Catanzaro et al. (2022) used the line list from the paper by
Castelli and Hubrig (2004) with atomic parameters calculated by
R. Kurucz\footnote{\texttt{ http://kurucz.harvard.edu/linelists/}}.
The line lists on the Kurucz' site have been updated in April 13, 2013 and now it is impossible
to find the difference between our $gf$ values and those from the work by Castelli and Hubrig (2004).
Therefore for 13 stars from Fossati et al. (2007) and Netopil et al. (2014) we used the published LTE
abundances without corrections.
Folsom et al. (2010, 2013) investigated three Am stars that are components of spectroscopic binaries
SB2.
Elemental abundances of these stars were determined with the method of synthetic spectra.
Atomic parameters of lines were taken from VALD.
Therefore we use the correction $-0.02$~dex for the LTE abundance of Sc bearing in mind the
difference in $gf$--values between Lawler et al. (2019) and Lawler and Dakin (1989).

Ca and Sc NLTE abundances for stars studied by Fossati et al. (2007), Netopil et al. (2014),
Niemczura et al. (2015), Catanzaro et al. (2022) and Folsom et al. (2010, 2013) we calculated
using the mean values of $\Delta_\mathrm{NLTE}$ similar to the case of Gebran+.

For the star HD~95608 (60 Leo) Adelman et al. (1999) obtained
$\log N_\mathrm{Sc}/N_\mathrm{A} = -10.93$
using the only line Sc~II 4246\AA\ (EW=4 m\AA).
We doubted that this line can be correctly measured in the star with $T_\mathrm{eff}=9250$\:K and
with so low Sc abundance.
For revision of the Ca and Sc abundances we used the spectrum from the archive UVES/VLT2
(program ID 0102.C-0547(A)).
To this end we used the values $T_\mathrm{eff}=9250$\:K, $\log g = 4.25$ and [Fe/H = 0.5
from Adelman et al. (1999), $gf$ values from Lawler et al. (2019), the stellar atmosphere models
were retrieved from the R. Kurucz' site\footnote{\tt http://kurucz.harvard.edu/grids/gridp05odfnew/}.
Calculations were done with the computer programs synthV\_NLTE (Tsymbal et al. 2019) and
BinMag (Kochukhov 2018).
We found that the line Sc~II 4246\AA\ cannot be extracted from the noise but one can evaluate
the upper abundance limit using the line Sc~II 3613\AA: $\log N_\mathrm{Sc}/N_\mathrm{A}\le-10.74$
($\log gf=0.42$, LTE), that is $[\textrm{Sc}/\textrm{H}]\le -1.64$ (NLTE).
The Ca abundance is determined using the lines Ca~I 4226, 4302\AA\ and Ca~II 3933\AA.

As seen in Figs.~\ref{fig2} and \ref{fig3}, the star 60~Leo is characterized by deficiency of
Ca typical for Am stars of similar
temperatures but at the same time has an extremely
low Sc abundance.
LeBlanc et al. (2015) investigated the high--resolution spectra and reported on existence of
vertical stratification of iron in the atmosphere of 60~Leo so that this star should belong to
Ap stars despite the absence of perceptible magnetic field.
We think that the conclusion on iron stratification should be checked using the NLTE analysis of
iron lines and leave 60~Leo in our sample of Am stars though in following discussion we
consider this star as an exception.

\begin{figure}
	\centering
	\includegraphics[width=0.45\columnwidth,clip]{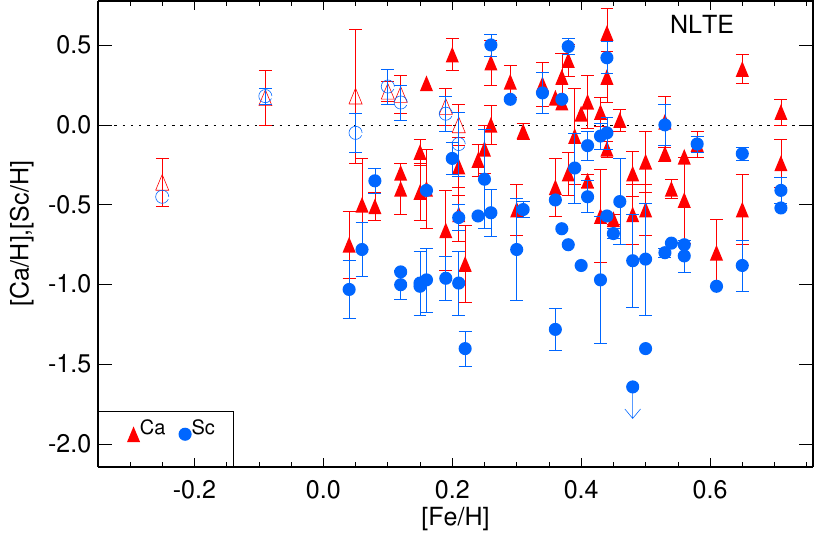}
    \includegraphics[width=0.45\columnwidth,clip]{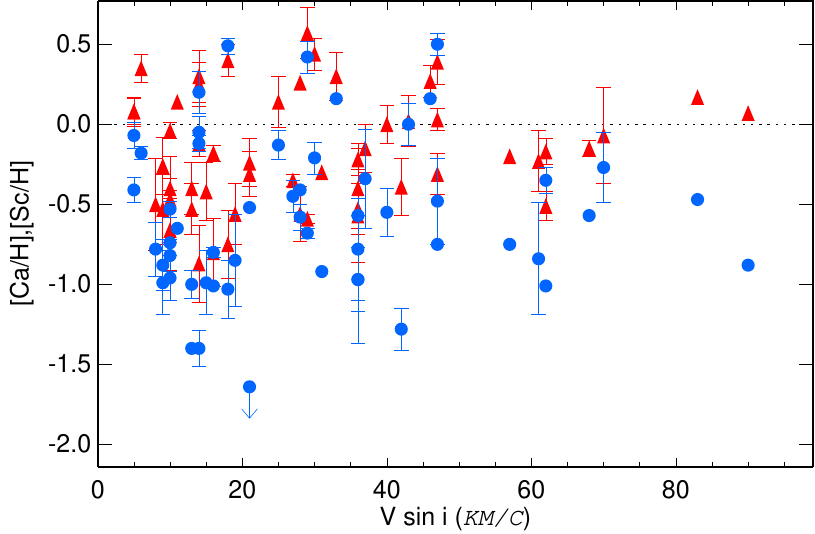}
	\caption{NLTE abundances [Ca/H] (triangles) and [Sc/H] (circles) in Am stars
	(filled symbols) and superficially normal stars (open symbols in the left panel) against [Fe/H] and $V \sin i$.}
	\label{fig2}
\end{figure}

\begin{figure}
	\centering
	\includegraphics[width=0.45\columnwidth,clip]{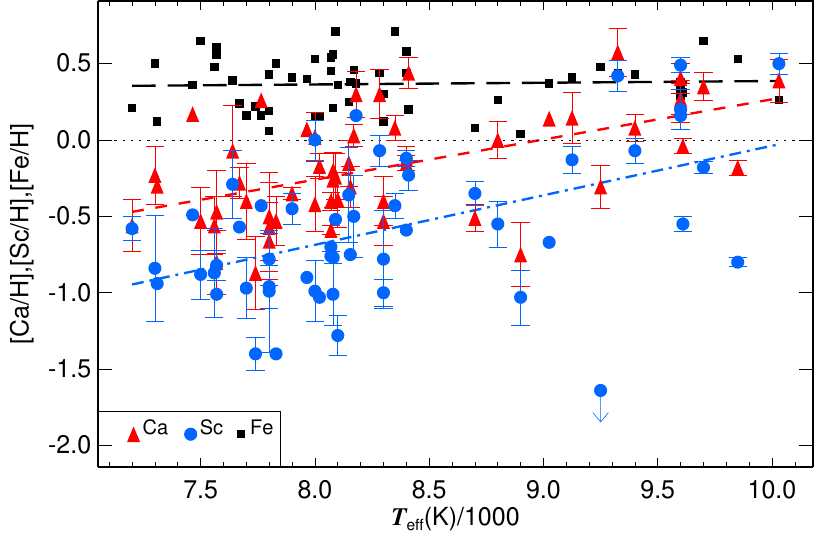}
	\includegraphics[width=0.45\columnwidth,clip]{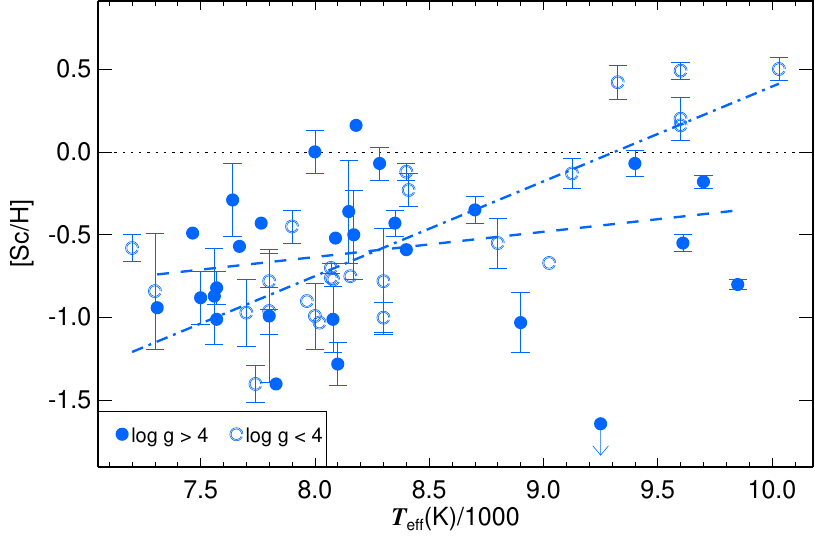}
	\caption{Left panel: NLTE abundances [Ca/H] (triangles), [Sc/H] (circles) and [Fe/H] (squares)
	in Am stars versus $T_\mathrm{eff}$.
	Right panel: [Sc/H] versus $T_\mathrm{eff}$ for $\log g \ge 4$ (filled circles) and
    $\log g < 4$ (open circles).
    The straight lines represent the linear fits.}
	\label{fig3}
\end{figure}

The LTE and NLTE absolute abundances $\log\varepsilon$ and NLTE abundance ratios [X/H] for
Ca and Sc are listed in Table~\ref{table1} and are shown in Figs.~\ref{fig2} and \ref{fig3}
(only NLTE).
The root mean square (r.m.s.) error is calculated as a standard deviation
$\sigma = \sqrt{\sum(x-\bar{x})^2/(N_l - 1)}$, where $N_i$ is the number of lines.
The value $\sigma=0$ implies that the source of data provides the mean abundance without its
error or the Sc abundance was evaluated only for one line.

The most conspicuous feature in Fig.~\ref{fig2} is the star HD~189849 (15~Vul) with deficiency
of Fe ([Fe/H]=-0.25), Ca and Sc.
Adelman et al. (1997) classified HD~189849 as a superficially normal A star.
This star was obviously included by mistake in the catalogue of Gazaryan et al. (2018).
Despite the significant deficiency of Sc in HD~58142 (21~Lyn, [Sc/H] < -0.3, LTE) this star was
classified as a superficially normal A star (Royer et al. 2014).
Our NLTE calculations allowed us to eliminate the Sc deficiency and corroborated the status of
HD~189849 as the superficially normal A star.
Two more stars have been earlier classified as superficially normal A stars: HD~43378
(2~Lyn, Caliskan and Adelman, 1997) and HD~72942 (Fossati, 2007) and results of our NLTE calculations
confirm those conclusions.
NLTE abundances of Ca and Sc as well as the metallicity [Fe/H] in three stars from the catalogue
by Gazaryan (2018) agree within errors with their solar counterparts so that in our study we
considered them as superficially normal A stars.
These are HD~34364B, HD~97633 and HD~109307.
Seven superficially normal A stars are listed in Table~\ref{table2}.
Therefore, our sample of Am stars shrinks to 54 stars (see Table~\ref{table1}) and
it is these stars that are discussed below from the point of view of their chemical peculiarity.

\begin{table}
 \renewcommand{\arraystretch}{1.0}
	\renewcommand{\tabcolsep}{3pt}
	\caption{Superficially normal A stars: their LTE and NLTE abundances of Ca and Sc}
	\vspace{3mm}
	\label{table2}
    \begin{tabular}{rrcrccrccrrcc}
\hline \noalign{\smallskip}
\multicolumn{1}{l}{HD} & \multicolumn{1}{c}{$T_\mathrm{eff}$} & $\log g$ & [Fe/ & \multicolumn{2}{c}{$\varepsilon$(Ca)} & [Ca/ & \multicolumn{2}{c}{$\varepsilon$(Sc)} & [Sc/ & [Ca/ & $V\sin i$ & Ref \\
\cline{5-6}
\cline{8-9}
    &  \multicolumn{1}{c}{(K)}  &      &  H]  & LTE & NLTE & H] & LTE & NLTE & H] & Sc] & (km/s) & \\
    	\noalign{\smallskip} \hline \noalign{\smallskip}
34364B & 10350 & 4.28 &   0.21 &  5.99 &  6.27\scriptsize{(0.13)} & --0.00 &  2.52 &  2.90\scriptsize{(0.20)} & --0.14 &   0.14 &  23 & 1  \\
43378  &  9295 & 4.10 & --0.09 &  6.37 &  6.44\scriptsize{(0.17)} &   0.17 &  3.08 &  3.22\scriptsize{(0.05)} &   0.18 & --0.01 &  46 & 2  \\
58142  &  9500 & 3.75 &   0.05 &  6.30 &  6.45\scriptsize{(0.42)} &   0.18 &  2.68 &  2.97\scriptsize{(0.12)} & --0.07 &   0.25 &  19 & 3  \\
97633  &  9330 & 3.66 &   0.19 &  6.31 &  6.38\scriptsize{(0.12)} &   0.11 &  2.85 &  3.11\scriptsize{(0.11)} &   0.07 &   0.04 &  23 & 4  \\
72942  &  8450 & 3.90 &   0.12 &  6.48 &  6.46\scriptsize{(0.12)} &   0.19 &  3.14 &  3.16\scriptsize{(0.11)} &   0.12 &   0.07 &  70 & 5  \\
109307 &  8396 & 4.10 &   0.10 &  6.50 &  6.48\scriptsize{(0.06)} &   0.21 &  3.25 &  3.26\scriptsize{(0.11)} &   0.22 & --0.01 &  14 & 6  \\
189849 &  7850 & 3.70 & --0.25 &  5.93 &  5.91\scriptsize{(0.15)} & --0.36 &  2.61 &  2.59\scriptsize{(0.12)} & --0.45 &   0.09 &  13 & 7  \\
    	\noalign{\smallskip}\hline \noalign{\smallskip}
\end{tabular}
\\
\textbf{Remarks.}
The numbers in parentheses are the r.m.s. errors $\sigma$.
References:
1 -- Folsom et al. (2010),
2 -- Caliskan and Adelman (1997),
3 -- Royer et al. (2014),
4 -- Adelman et al. (2015),
5 -- Fossati et al. (2007),
6 -- Fossati et al. (2008),
7 -- Adelman et al. (1997).

\end{table}

\section{Dependence of chemical peculiarity on physical parameters of Am stars}

Am stars of our sample exhibit in their atmospheres the iron excess compared to the solar
abundance ($0.06\le [\textrm{Fe}/\textrm{H}]\le 0.71$) and at the same time the Ca and Sc abundances
varying in wide ranges: $-0.87\le[\textrm{Ca}/\textrm{H}]\le +0.57$ and
$-1.64\le[\textrm{Sc}/\textrm{H}]\le +0.50$.
Why the chemical peculiarity varies within so wide range?
To answer to this question we carried out the statistical analysis of the Ca and Sc abundance data
in order to find their possible correlations with physical parameters of the stars.
There is no correlation between the Ca or Sc abundances and [Fe/H] or $V\sin i$ (see Fig.~\ref{fig2}).
However, as seen in Fig.~\ref{fig3}, deviations in the abundances of Ca and Sc from those of the Sun
decrease with increasing $T_\mathrm{eff}$.
Moreover, the linear fits of [Ca/H] and [Sc/H] with respect to $T_\mathrm{eff}$ have the larger slope
for stars with $\log g<4$ than that for $\log g \ge 4$ (see the plot for [Sc/H] in the left panel
of Fig.~\ref{fig3}), so that deficiencies in Ca and Sc change to their excesses for
$T_\mathrm{eff}>9300$\:K and $\log g<4$.
The iron excess is on average independent of the effective temperature.

Following Gazaryan et al. (2018) we calculated the Spearman’s rank correlation coefficients $\rho$
in order to test the existence of dependencies between the Ca or Sc abundances on the one hand
and $T_\mathrm{eff}$, [Fe/H], $V\sin i$ on the other hand without any assumptions on the shape
of the dependence.
The calculations were carried out for the full sample of 54 stars.
Dependence of [Ca/H] and [Sc/H] on $T_\mathrm{eff}$ is confirmed by relatively high values of
$\rho(\textrm{Ca}-T_\mathrm{eff})=0.54$ and $\rho(\textrm{Sc}-T_\mathrm{eff})=0.46$.
If we exclude 60~Leo then $\rho(\textrm{Sc}-T_\mathrm{eff})=0.52$.
In the tests of the abundance dependence on [Fe/H] we obtained $\rho(\textrm{Ca}-\textrm{Fe})=0.10$
and $\rho(\textrm{Sc}-\textrm{Fe})=0.14$ whereas the tests of their dependence on $V\sin i$ gave
$\rho(\textrm{Ca}-V\sin i)=0.17$ and $\rho(\textrm{Sc}-V\sin i)=0.12$.
Rotation axes of stars have no specific direction in the space so that we can assert that
Ca and Sc abundances are independent not only of the velocity projection on the line--of--sight
but also of the actual stellar rotation velocity.

Gazaryan et al. (2018) searched for correlations between abundances (Ca and Sc) and fundamental
parameters ($T_\mathrm{eff}$, $\log g$ and $V\sin i$) for the sample of 105 AmFm stars but they found
the only correlation between [Ca/H] and $T_\mathrm{eff}$ with significantly smaller coefficient
$\rho(\textrm{Ca}-T_\mathrm{eff}) = 0.28$ in comparison with that given above.
This seems due to the fact that Gazaryan et al. (2018) used less homogeneous data and did not take into
account NLTE effects.
The same reasons seem to be responsible for their failed search for correlation between Sc and
$T_\mathrm{eff}$ in the sample consisting of 64 stars: $\rho(\textrm{Sc}-T_\mathrm{eff}) = -0.03$.

\begin{figure}
    \centering
	\includegraphics[width=0.45\columnwidth,clip]{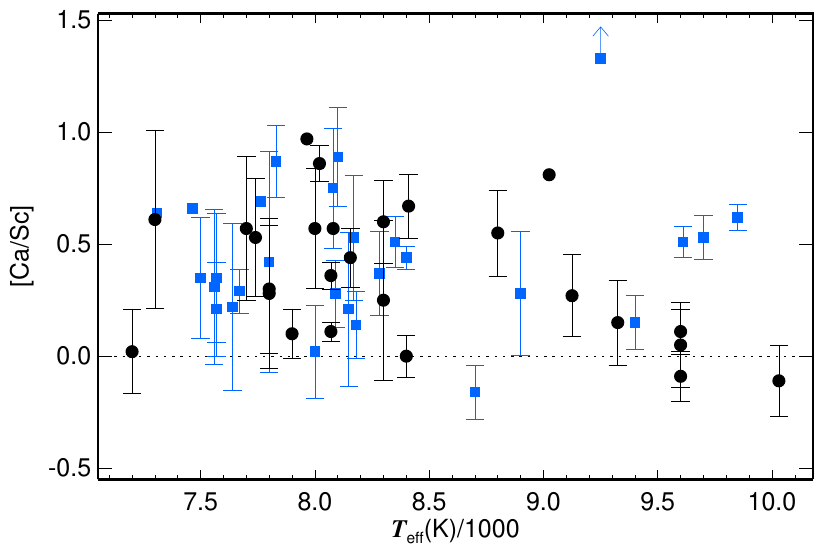}
	\caption{NLTE ratios [Ca/Sc] in Am stars against $T_\mathrm{eff}$.
	Squares and circles represent the stars with $\log g \ge 4$ and $\log g < 4$, respectively.}
	\label{fig4}
\end{figure}

Fig.~\ref{fig4} shows the abundance ratios [Ca/Sc] for stars with $\log g \ge 4$ and $\log g < 4$.
Mean values of [Ca/Sc] for the full sample (54 stars), the sample for 28 stars with $\log g\ge 4$ and
the sample for 26 stars with $\log g < 4$ agree with each other: $0.41\pm 0.30$, $0.44\pm 0.30$ and
$0.37\pm 0.30$, respectively, though there is the large scatter among stars with close values of
$T_\mathrm{eff}$, [Fe/H] and $V\sin i$.
Our statistics is fairly small but one can assume that in stars with high effective temperatures
($T_\mathrm{eff}\gtrsim 9500$\:K) the atomic diffusion processes are different depending on $\log g$
and, therefore, the star age.
Mashonkina (2024) noted the existence of the roughly constant abundance ratio
$[\textrm{Ca}/\textrm{Sc}]\simeq 0.6-0.7$ in Am stars and concluded that the large value of [Ca/Sc]
is a feature of Am stars.
We should recognize that this is not true because the sample in the work of Mashonkina (2024) consisted of
15 Am stars mostly with $T_\mathrm{eff}<9500$\:K with only three hotter stars with $\log g\le 4.1$.
In the more extended sample we found the stars with the iron excess and
$[\textrm{Ca}/\textrm{H}]\simeq [\textrm{Sc}/\textrm{H}]$, that is with [Ca/Sc] close to zero.

\section{Comparison with diffusion models of main sequence stars}

The superficial element abundances change with the evolution time so that to compare observations with
diffusion models we selected the Am stars that are members of open clusters of known ages.
These are the Pleiades, Praesepe, Hyades and Coma clusters.
Ages and metallicities $[\textrm{Fe}/\textrm{H}]_0$ of these open clusters were derived
by Netopil et al. (2022) and are listed in Table~\ref{table1}.
We assumed that the initial abundances of these clusters satisfy the condition
$[\textrm{Ca}/\textrm{H}]_0 = [\textrm{Sc}/\textrm{H}]_0$ = $[\textrm{Fe}/\textrm{H}]_0$.

\subsection{Diffusion models in published works}

Stellar evolution calculations including the effects of atomic diffusion in stars with masses from
1.45 to $3.0M_\odot$ (Richer et al. 2000) exhibit iron accumulation in layers with temperature
$\sim 2\times 10^5$\:K (the Z--bump) where the more opaque gas is unstable against convection and
undergoes mixing as it occurs in hydrogen and helium ionization zones.
During evolution these three mixing zones join into a surface mixing zone (SMZ).
Therefore any change of the superficial abundance of elements in comparison with the initial abundance
is due to the processes that take place below the mixing zone.
The diffusion models taking into account convective mixing predict growth of the superficial
abundance of Fe and at the same time decrease in Ca and Sc abundances in comparison with their initial
values but in the theory these effects are significantly larger than those observed in Am stars.
To decrease the superficial iron excess Richer et al. (2000) assumed the existence of additional
mixing in SMZ.
The authors called this phenomenon as turbulence which in one--dimensional geometry is considered as
turbulent diffusion with diffusion coefficient $\omega$ treated as a free parameter.

Vick et al. (2010) investigated the role of mass loss which also affects superficial abundances
in the diffusion stellar models.

\begin{figure*}
	\centering
    \begin{center}
    \begin{picture}(400,260)
	\put(-30,150){\includegraphics[width=0.45\columnwidth,clip]{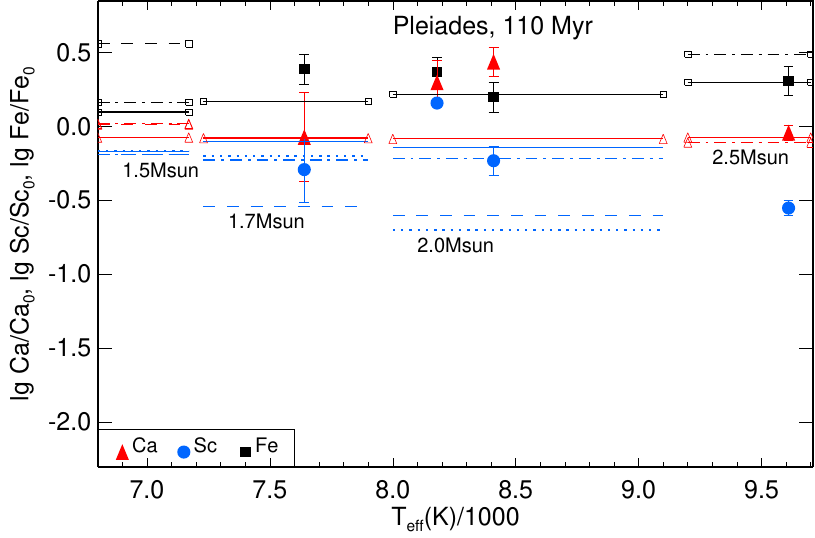}}
	\put(210,150){\includegraphics[width=0.45\columnwidth,clip]{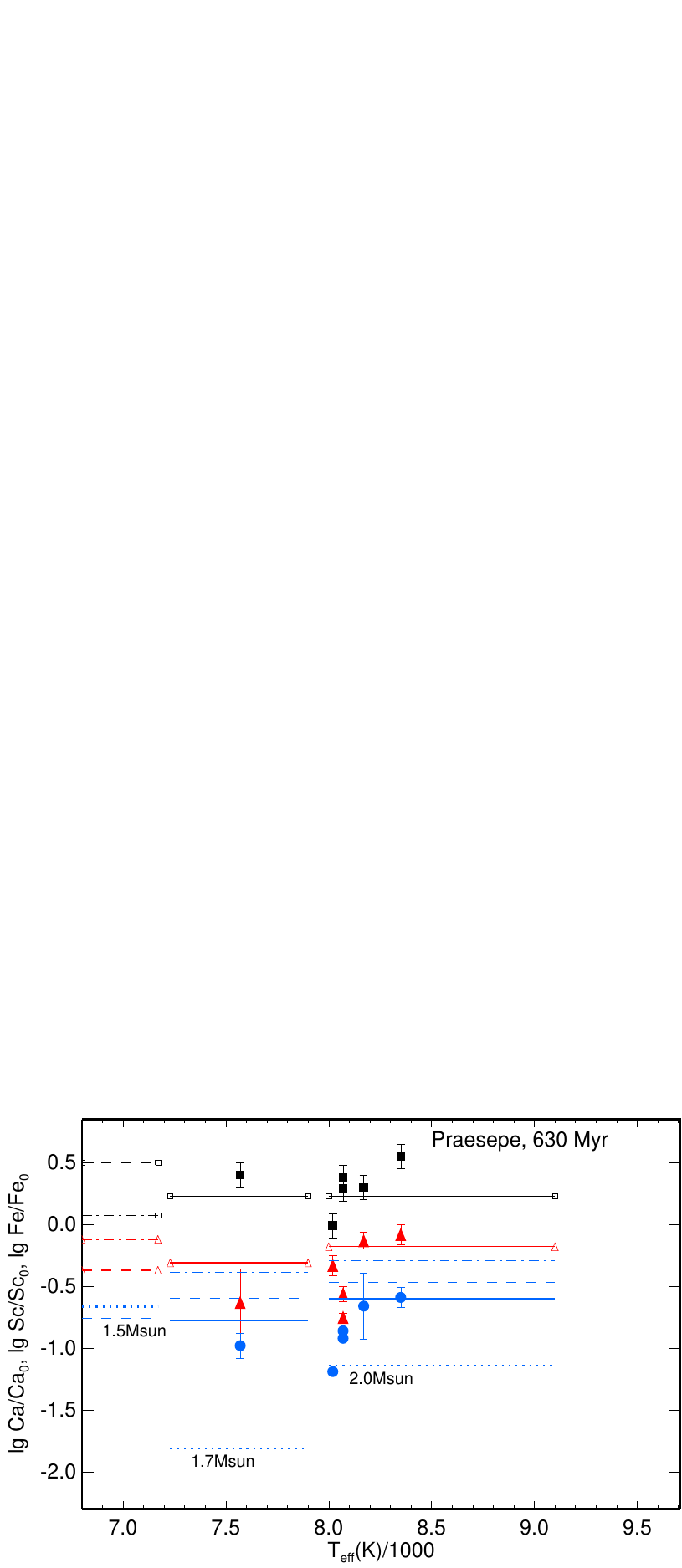}}
	\put(-30,  0){\includegraphics[width=0.45\columnwidth,clip]{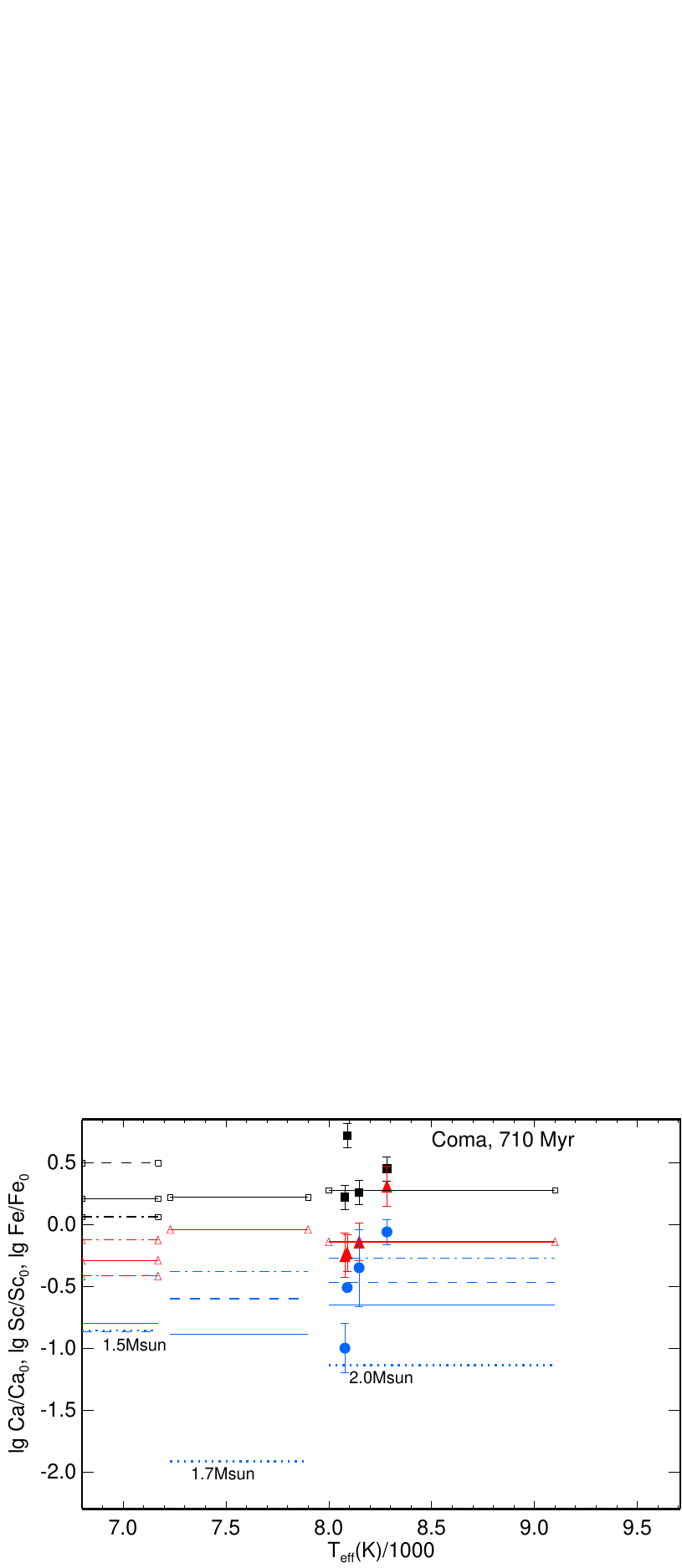}}
    \put(210,  0){\includegraphics[width=0.45\columnwidth,clip]{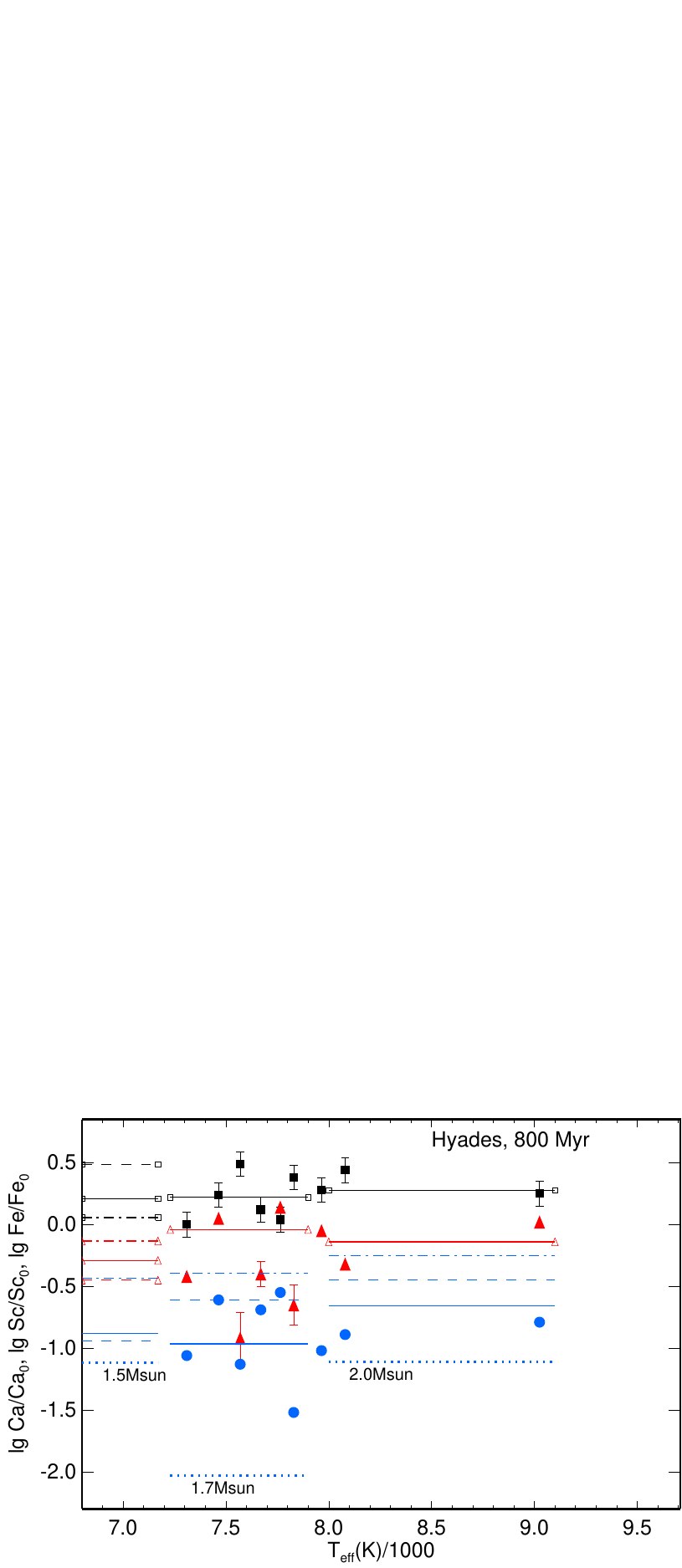}}
    \end{picture}
    \end{center}
	\vspace{-4mm}
	\caption{Observed variations of the superficial abundances of Ca (triangles), Sc (circles) and Fe (squares)
    with respect to their initial values in Am stars of open clusters and the diffusion models with different masses.
	Solid lines -- models R1K-2 (Ca, Fe) and RMT500 (Sc);
	dotted lines -- RMT50 (Sc);
	dashed lines -- ML2m14 (Ca, Fe) and MLm14 (Sc);
	dashed--dotted lines -- MLm13(Ca, Fe, Sc)
	For designations of diffusion models see the text.}
	\label{fig5}
\end{figure*}

Fig.~\ref{fig5} shows the Ca, Sc and Fe abundances observed in Am stars of the four open clusters.
The abundances are plotted with respect to their initial values and are compared with the superficial
abundances of the diffusion models.
For Ca and Fe these are the turbulent models R1K-2 (see Figs.~12, 14 and 16 of the paper by
Richer et al. 2000) computed with the turbulent diffusion  coefficient $\omega=10^3$.
The mass of the superficial turbulent mixing zone is $\log(1-M_r/M)\simeq-4.6$, where $M_r$ is the
Lagrangian mass coordinate and $M$ is the stellar mass.
Similar models for Sc were computed by Hui--Bon--Hoa et al. (2022) but with $\omega=50$ (model RMT50) and
$\omega=500$ (model RMT500, see Fig.~6 of their paper).
We consider also the diffusion models computed with effects of mass loss by Vick et al. (2010) for the
mass loss rate $\dot M = 10^{-13}M_\odot/\textrm{yr}$ (model MLm13) and
$\dot M = 2\times 10^{-14}M_\odot/\textrm{yr}$ (model ML2m14) as well as by Hui--Bon--Hoa et al. (2022)
for Sc with $\dot M = 10^{-13}M_\odot/\textrm{yr}$ (model MLm13) and
$\dot M = 10^{-14}M_\odot/\textrm{yr}$ (model MLm14).

The diffusion models correspond to certain values of the stellar mass and the star age.
Stellar masses in the open clusters are not known and to compare the theory with observations we
use the relation $M-T_\mathrm{eff}$ for the main sequence stars (Cox 2000).
The plots in Fig.~\ref{fig5} allow us to conclude the following.

\begin{itemize}
 \item For Fe the turbulent models (R1K-2) weakly depend on the stellar mass and  fairly well
 reproduce observations at star ages from 110 to 800 Myr.
 \item For Ca and Sc the turbulent models are almost independent of the stellar mass
 as long as the star is still young (110 Myr).
 In models of older stars the Sc deficiency decreases with increasing stellar mass $M$
 because $T_\mathrm{eff}$ of the zero--age stars depends on $M$.
 The same effect was found for Ca in the models with age 630 Myr and the opposite effect for models
 of older stars.
 \item Diffusion models taking into account the mass loss can be compared with observations only for
 Fe and Ca in the case of $M=2.5M_\odot$ and the age 110 Myr.
 For Ca this model as well as the turbulent models provide the close results and reproduce
 observations of the only star available.
 For Fe the turbulent model seems to be more preferable than the model MLm13.
 \item
 Models with mass loss can be compared with turbulent models only for Sc.
 The model with relatively weak turbulence RMT50 predicts too large deficiency of Sc  which is
 not observed in the stars.
 Observations of youngest stars (110 Myr) are fairly well reproduced by the turbulent models RMT500
 as well as by the model with high mass loss rate MLm13.
 Mass loss in older stellar models leads to the smaller Sc deficiency in comparison with turbulent
 models so that the models RMT500 are in better agreement with observations of the stars in
 Praesepe and Hyades clusters than the models MLm13.
 Bearing in mind the scatter of observational data we can conclude that diffusion models with low
 mass loss rate MLm14 also agree with observations.
\end{itemize}

In general we conclude that the turbulent models with large values of the turbulence parameter
($\omega=10^3$ for Ca and Fe) and ($\omega=500$ for Sc) fairly well reproduce observations of
Ca, Sc and Fe abundances in Am stars of open clusters.
At the same time the diffusion models with mass loss $\dot M\sim 10^{-14}M_\odot/\textrm{yr}$
cannot be excluded because in some cases observations of stars of the same age exhibit large
scatter.

\subsection{Results of diffusion model calculations}

For more detailed comparisons between NLTE abundances and diffusion models we carried out the stellar
evolution calculations of main sequence stars with masses $1.5M_\odot\le M\le 3M_\odot$ for
initial abundances of helium and heavier elements $Y=0.28$ and $Z=0.02$, respectively.
To this end we used the program \textsc{MESA} version r23.05.1 (Jermyn et al. 2019).
Energy generation rates and nucleosynthesis were calculated for the reaction grid comprising 19
isotopes from ${}^1$H to ${}^{59}$Cu.
Convective mixing was treated following the theory of B\"ohm--Vitense (1959) with the
mixing length to pressure scale height ratio $\alpha_\mathrm{MLT}=1.8$.
Extra mixing at convection zone boundaries was taken into account according to Herwig (2000) for
the overshooting parameter $f_\mathrm{ov}=0.016$.
Mass loss due to the stellar wind was assumed to be negligible during the main sequence stage
(i.e. for the central hydrogen abundance $X({}^1\textrm{H})_\mathrm{c} > 10^{-4}$).

Together with solution of the equations of stellar evolution we calculated the changes in radial
distributions of elemental abundances due to effects of atomic diffusion.
The current version of the \textsc{MESA} program implements the solution of the Burgers equations
(Burgers 1969) extended by radiative acceleration terms (Hu et al. 2011).
Radiative accelerations were computed using the data base of monochromatic opacities for 17 elements
from hydrogen to nickel (Seaton 2005).
In the present study we solved the Burgers equations for all the 19 elements of the nuclear reaction
grid whereas effects of radiative acceleration were considered for atoms and ions of the
following 14 elements:
${}^1\textrm{H}$, ${}^4\textrm{He}$, ${}^{12}\textrm{C}$, ${}^{14}\textrm{N}$, ${}^{16}\textrm{O}$,
${}^{20}\textrm{Ne}$, ${}^{23}\textrm{Na}$, ${}^{24}\textrm{Mg}$, ${}^{27}\textrm{Al}$, ${}^{28}\textrm{Si}$,
${}^{32}\textrm{S}$, ${}^{40}\textrm{Ca}$, ${}^{56}\textrm{Fe}$ and ${}^{58}\textrm{Ni}$.

\begin{figure}
\begin{center}
\includegraphics[width=0.45\textwidth]{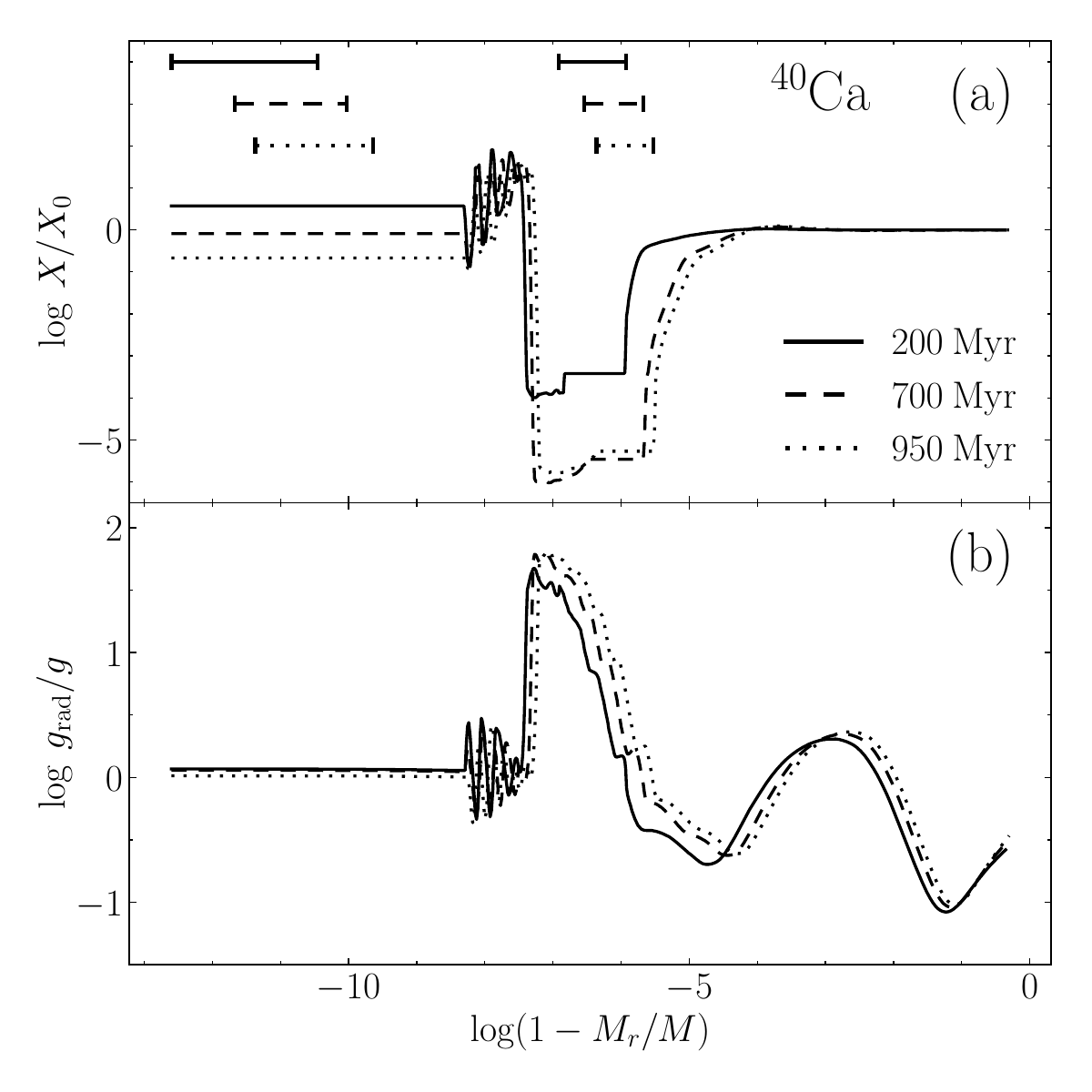}
\includegraphics[width=0.45\textwidth]{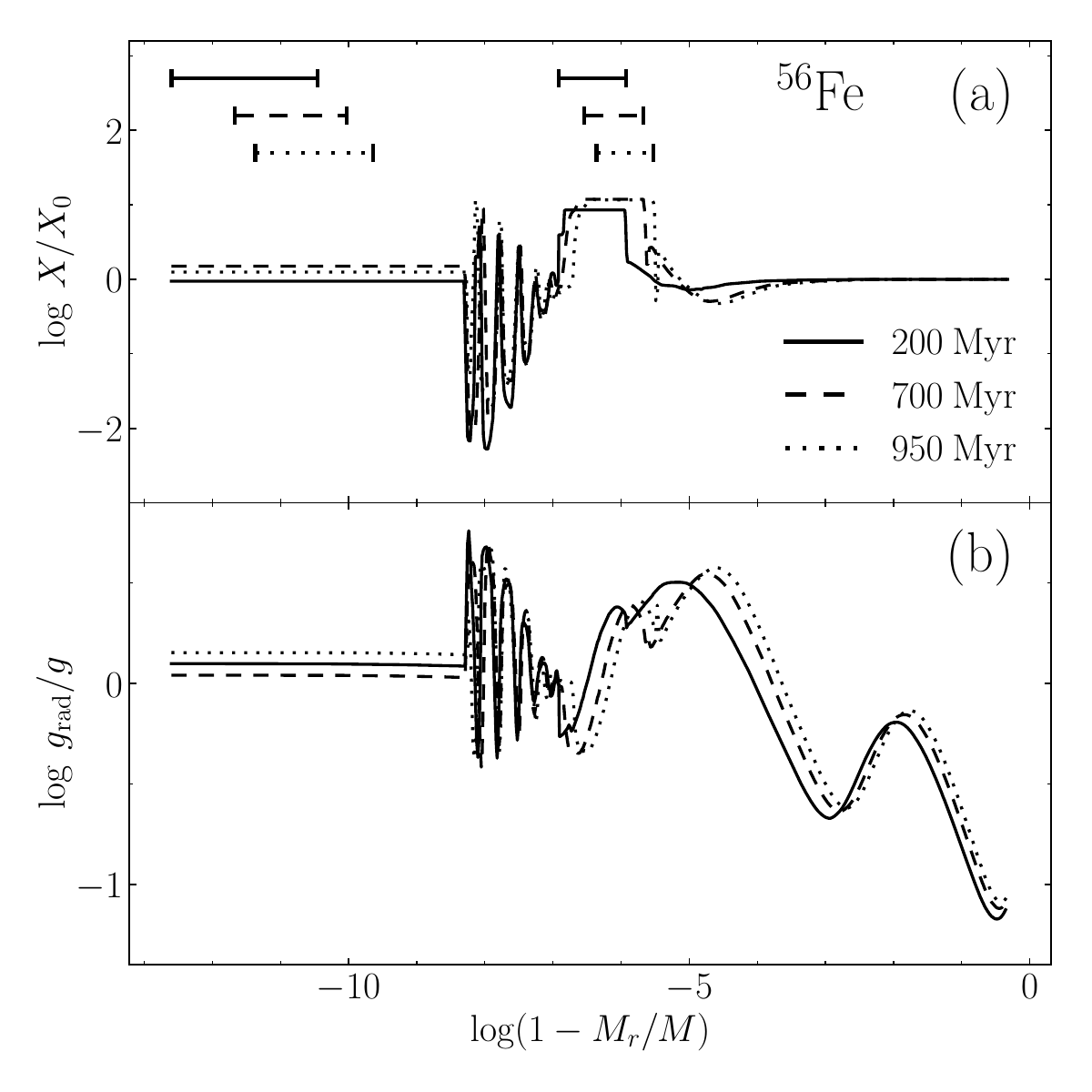}
\end{center}
\caption{The ratios of Ca (left panel) and Fe (right panel) abundances to their initial values (a)
    and the ratios of radiative acceleration on atoms of Ca (left panel) and Fe (right panel) to the
    gravitational acceleration (b) as a function of the Lagrangian mass coordinate for the models of
    the evolutionary sequence $M=1.7M_\odot$ with the age 200 Myr (solid lines), 700 Myr (dashed lines) and
	950 Myr (dotted lines). The ranges of convection zones are depicted at the upper part of the plots.}
	\label{fig6}
\end{figure}

Results of our calculations show that excess or deficiency in abundance of any element depends on
the location of the maximum of radiative acceleration acting on atoms of this element with respect to
the boundary of the convection zone.
This is illustrated in Fig.~\ref{fig6} where the radial distributions of Ca and Fe abundances
are plotted for the star with mass $M=1.7M_\odot$ at three values of the star age: $t_\mathrm{ev}=200$,
700 and 950 Myr.
There are two convection zones in the stellar envelope.
The outer convection zone comprises the layers of hydrogen and helium ionization
($10^4\:\textrm{K}\lesssim T \lesssim 3\times 10^4\:\textrm{K}$)
whereas the inner convection zone locates in layers with temperature
$10^5\:\textrm{K}\lesssim T \lesssim 3\times 10^5\:\textrm{K}$
and is due to opacity maximum by elements of the iron group.
The bounds of the convection zones are depicted in the upper part of Fig.~\ref{fig6}.

In the layers with $g_\mathrm{rad} < g$ ($g=GM_r/r^2$ is the gravitational acceleration in the layer
with the radius $r$ and Lagrangian mass coordinate $M_r$) the Ca abundance decreases with time
(see Fig.~\ref{fig6}) because atoms of calcium are ionized to a Ne--like state Ca~XI
(see Fig.~\ref{fig7}) and are inefficient absorbers of radiation so that  gravitational settling leads
to significant reduction of Ca abundance in these layers.
Near the outer boundary of the inner convection zone the ionization of Ca atoms changes
so that increasing absorption by ions of calcium becomes responsible for their radiative acceleration.
The abundance of calcium in the convective mixing zone decreases with time because its atoms
are pushed out by radiative pressure beyond the convection zone.

\begin{figure*}
	\centering
	\hspace{-4mm}
	\includegraphics[width=0.50\columnwidth,clip]{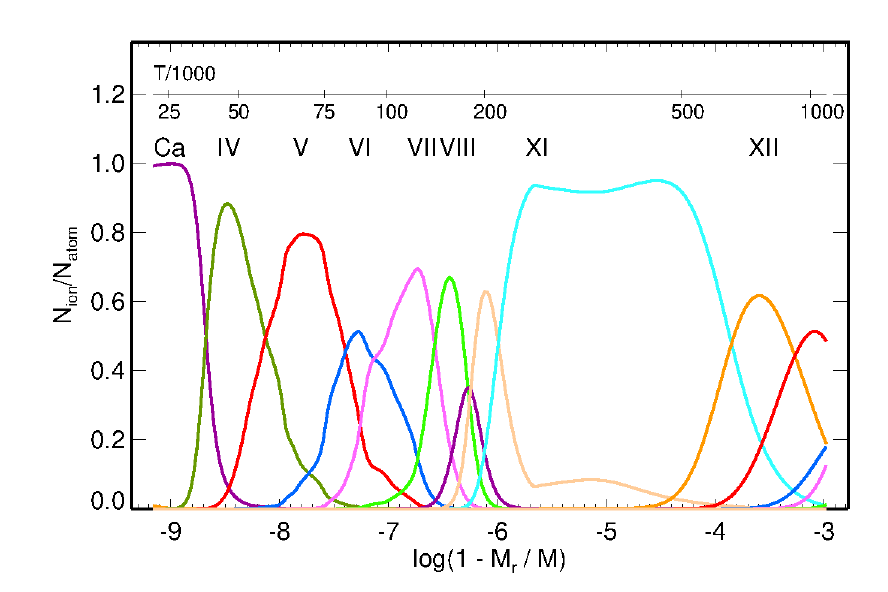}
	\hspace{-5mm}
	\includegraphics[width=0.50\columnwidth,clip]{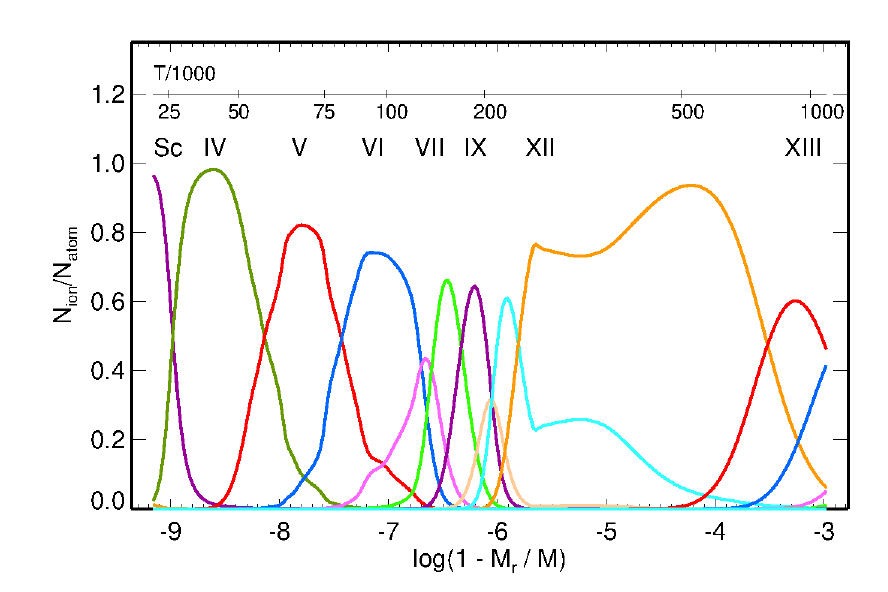}
	\caption{The ratios of Ca and Sc ion concentration to the full concentration of atoms of these
	elements as a function of the depth $\log (1 - M_r/M)$ for the model with mass $M = 1.7 M_\odot$
	and the age 700 Myr.}
	\label{fig7}
\end{figure*}

The region of the Ne--like state of Fe atoms locates substantially deeper than the inner convection zone
whereas the opacity maximum (Z bump) is due to numerous absorption lines of most abundant ions Fe~XIII
locating a little deeper than the inner convection zone.
The iron atoms are pushed out by radiative pressure toward the inner convection zone and this is the cause
of significant excess of the iron abundance in the layers with temperature $T\sim 2\times 10^5$\:K.

The absence of data on monochromatic opacity of Sc in the data base of the Opacity Project (Seaton 2005)
did not allow us to calculate evolutionary changes of Sc superficial abundances.
However, as seen in Fig.~\ref{fig7}, the ionization state of Sc changes with depth similar to that
of Ca.
Dominant Ne--like ions Sc~XII within the layers with temperature
$2.5\times 10^5\:\textrm{K} < T < 5\times 10^5\:\textrm{K}$
should be responsible for decrease of Sc abundance in these layers and should lead to deficiency of Sc
superficial abundances similar to Ca.

\begin{figure}
    \centering
	\includegraphics[width=0.70\columnwidth,clip]{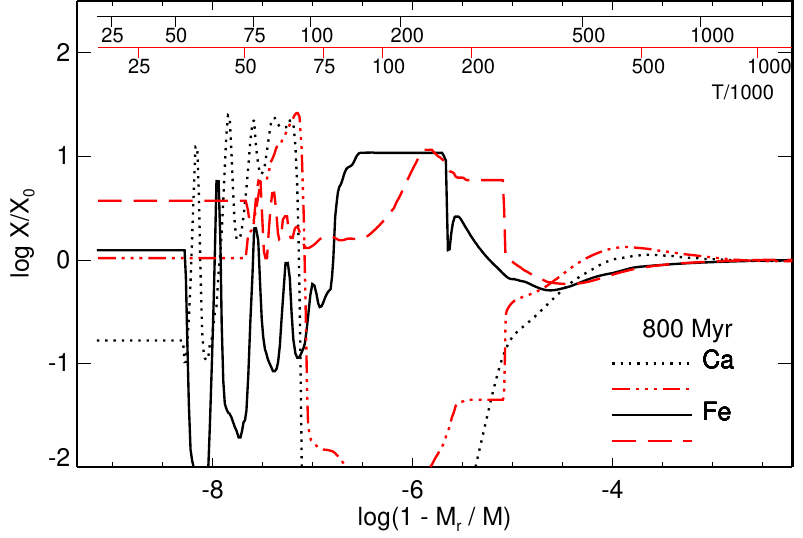}
	\caption{The ratios of Ca and Fe abundances to their initial values as a function of the depth
	$\log (1 - M_r/M)$ in models with mass $M=1.5$ and $2.2M_\odot$ at the age 800 Myr.
	The initial chemical abundance is assumed to be the solar one.
    The dotted and dashed--three dotted lines correspond to Ca in the $M=1.5$ and $2.2M_\odot$ models,
    respectively, whereas Fe in the same models is shown by the solid and dashed lines.}
	\label{fig8}
\end{figure}

Fig.~\ref{fig8} illustrates dependence of Ca and Fe abundance profiles on the stellar mass for
$M=1.5$ and $2.2M_\odot$ and the star age 800 Myr.
Element separation due to atomic diffusion takes place in the outer layers with mass $\sim 10^{-3}M$.
Iron is accumulated in the layers with $1.5\times 10^5\:\textrm{K} < T < 2.5\times 10^5\:\textrm{K}$.
Both models exhibit excess in the iron abundance in comparison with its initial value but the Fe
abundance excess is higher in the star with the larger mass.
Abundance of Ca drops abruptly within the Lagrangian coordinates $-7 < \log(1 - M_r/M) < -5$ by six
and two orders of magnitude in models with $M=1.5$ and $2.2M_\odot$.
In the model $M=1.5M_\odot$ the superficial Ca abundance is significantly deficient
($\log\textrm{Ca}/\textrm{Ca}_0=-0.78$) whereas in the model with $M=2.2M_\odot$ the superficial
abundance of Ca is close to the solar abundance.

\textbf{Comparison with observations.}

\begin{figure*}
	\includegraphics[width=0.45\columnwidth,clip]{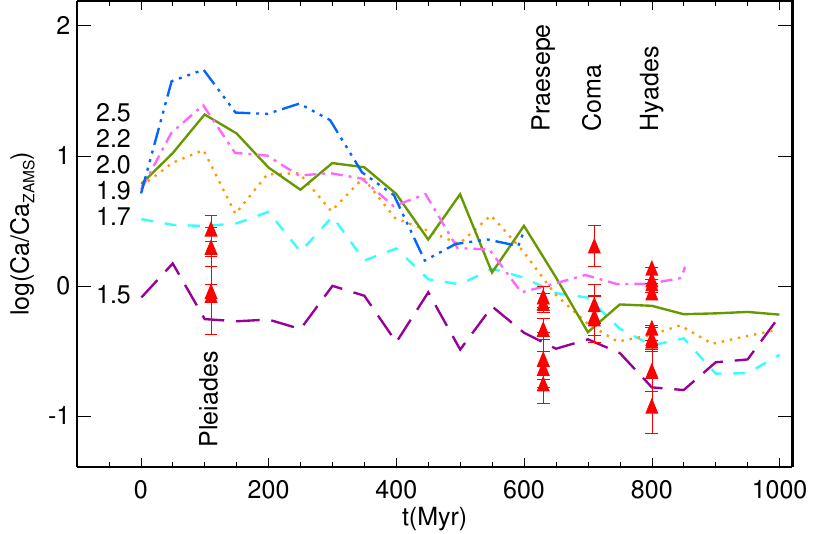}
	\includegraphics[width=0.45\columnwidth,clip]{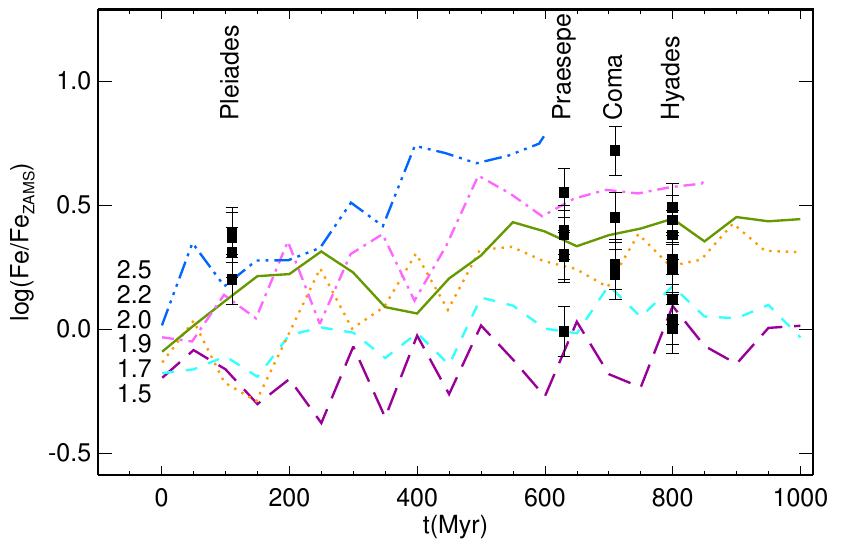}
	\caption{Observed abundances of Ca (left panel, triangles) and Fe (right panel, squares)
	in atmospheres of Am stars in units of their initial values of the parent open cluster
	compared to the superficial Ca and Fe abundances in the diffusion models computed with the
	\textsc{MESA} program for several values of the stellar mass $M$.
    Here $M=1.5M_\odot$ (long dashed line), 1.7 (dashed line), 1.9 (dotted line), 2.0 (solid line),
    2.2 (dash--dotted line) and 2.5 (dash--three dotted line).}
	\label{fig9}
\end{figure*}

Evolutionary changes in superficial abundances of Ca and Fe in diffusion models with masses from 1.5 to
$2.5M_\odot$ are shown in Fig.~\ref{fig9} where we plot also the observed abundances of Ca and Fe
 in the open clusters of different ages.
The masses of Am stars are unknown and therefore we can conclude that the theory agrees with observations.
However one should bear in mind that the age of the Pleiades cluster is only 110 Myr whereas the age of
other three clusters is greater than 600 Myr.
Ca and Fe abundances in Am stars of older clusters indicate almost the same mass ranges:
from 1.7 to $2.2M_\odot$ for the Coma cluster and from 1.5 to $2.2M_\odot$ for the Praesepe and
Hyades clusters.
However the Ca and Fe abundances in the Pleiades cluster can be explained only with different model masses:
$1.5M_\odot\le M\le  1.7M_\odot$ for Ca and $M>2M_\odot$ for Fe.

\subsection{Sirius: comparison with diffusion models}

Sirius is one of the well--studied Am stars.
But are the diffusion models able to reproduce its chemical composition?
Fig.~\ref{fig10} shows the Sirius abundances of 18 elements from He to Ni (Romanovskaya et al. 2023)
and the Sc abundance according to Mashonkina (2024).
Abundances of 13 elements were obtained from NLTE calculations whereas available abundances of
V, Cr, Mn, Co and Ni correspond to the LTE.
NLTE corrections for these five elements are expected to be small enough and close to the average difference
$\log\varepsilon_\mathrm{NLTE} - \log\varepsilon_\mathrm{LTE} = -0.01$~dex evaluated for Fe.

\begin{figure}
    \centering
	\includegraphics[width=0.55\columnwidth,clip]{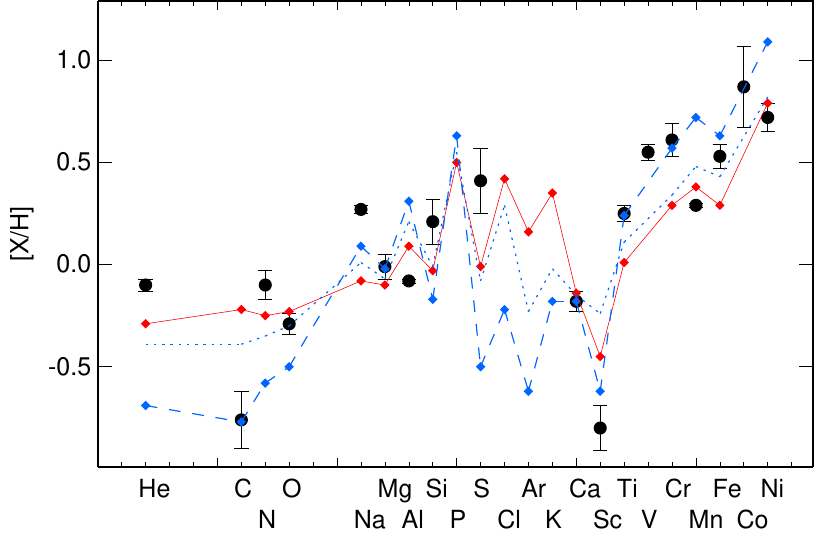}
	\caption{The elemental abundances in Sirius (circles) and predictions of the diffusion models:
	turbulent models R1K-2 and RMT500 (solid lines), the models with mass loss ML5m14 and MLm14
	(dashed line), the model with mass loss MLm13 (dotted line).}
	\label{fig10}
\end{figure}

The mass of Sirius is $M=2.143M_\odot$ and the mass of its white dwarf (WD) component was estimated
by Gatewood and Gatewood (1978).
Bearing in mind the duration of stellar evolution preceding the onset of the WD stage as well as
the cooling time of WD, Richer et al. (2000) evaluated the age of Sirius in the range from 200 to 300~Myr.
Therefore, for comparison with observations we considered the turbulent models R1K-2 with the mass $M=2.2M_\odot$
and age $t_\mathrm{ev}=300$ Myr (Richer et al. 2000) for abundances from He to Ni (with exception of Sc) and
the model RMT500 with $M=2.0M_\odot$, $t_\mathrm{ev}=300$ Myr (Hui-Bon-Hoa et al. 2022) for the
Sc abundance.
Moreover, we considered the models with mass loss ML5m14 and MLm13 (Vick et al. 2010) with $M=2.5M_\odot$,
$t_\mathrm{ev}=250$ Myr for abundances from He to Ni (except for Sc) and the models MLm14 and MLm13 with
the mass $M=2.0M_\odot$ and age $t_\mathrm{ev}=250$ Myr (Hui-Bon-Hoa et al. 2022) for the abundance
of Sc.
As seen in Fig.~\ref{fig10}, the turbulent models agree with the observations within $\pm 0.2$~dex
for a number of elements, such as
 He, N, O, Mg, Al, Si, Ca, Mn and Ni.
 A satisfactory agreement is obtained also for S and Co because of their large observational errors.
The diffusion models with mass loss also fail to reproduce the observed abundances of all elements
but, nevertheless, they seem to be more preferable for C, Na, Sc, Ti, V, Cr and Fe abundances.

The diffusion models computed in the present study with the code \textsc{MESA} for the mass $M=2M_\odot$
and age 250 Myr exhibit the superficial abundance ratios $\log\textrm{Ca}/\textrm{Ca}_0=0.74$ and
$\log\textrm{Fe}/\textrm{Fe}_0=0.31$ whereas the observed calcium abundance is significantly lower
($[\textrm{Ca}/\textrm{H}]=-0.18$) and the iron abundance is higher ($[\textrm{Fe}/\textrm{H}]= 0.53$).

The diffusion models of Richer et al. (2000) and Vick et al. (2010) have been earlier compared with
the observational data for Sirius in the studies by Richer et al. (2000), Landstreet (2011) and
Michaud et al. (2011).
Based on the LTE abundances of elements from He to Ni Landstreet (2011) and Michaud et al. (2011)
concluded that with exception of N and Na the diffusion models reproduce observations within $2\sigma$.
We should note the large error $\simeq 0.25$~dex in C abundance measurements (Landstreet 2011).
The authors of these works supposed that disagreement between observations and the diffusion models is due
to contamination of the atmosphere of Sirius by elements formed earlier as a result of nucleosynthesis in the
more massive component observed now as the white dwarf.

Our data on elemental abundances in the atmosphere of Sirius allow us to test the hypothesis on
affecting the C, N, O abundances by
mass transfer from the evolved component.
According to Landstreet (2011), the initial mass of the Sirius companion was $\sim 6M_\odot$.
During evolution this star passed through the phase of the dredge--up of the CNO cycle products with
superficial deficiency in C and excess in N.
It is this phenomenon that is observed in the atmospheres of B supergiants (Przybilla et al. 2010) but
according to the theory of stellar evolution the sum of abundances of the CNO cycle products should be
the same as that of their initial abundances.
In Sirius this sum is $\log\varepsilon(\textrm{C} + \textrm{N} + \textrm{O})=8.65$ and is smaller than
the solar value by 0.3~dex.
Therefore the C, N, O abundances in Sirius cannot be explained as a result of mass transfer from the
secondary component during its supergiant evolutionary phase.
To explain significant carbon deficiency in Sirius ([C/H] = $-0.69$ according to Mashonkina et al. 2020)
as a results of mass transfer during the AGB phase of the second component we have to assume that
during this phase carbon experienced its destruction rather than synthesis.
It should be noted also that abundances of s--elements (Sr, Y, Zr, Ba) in Sirius are not higher than
those in other Am stars (Romanovskaya et al. 2023).

The problem of Ca and Fe abundances in stars of the young stellar cluster Pleiades seems to be similar to
that of Sirius.
One of the causes of the disagreement is perhaps due to more complicated physics of element separation
processes that have not yet been accounted for by the stellar diffusion models.
For example, the star St1612 is a member of the open cluster Stock~16 with the age evaluated from 4 to 6 Myr
(Netopil et al. 2014), but at the same time this star exhibits explicit features of the Am star:
[Fe/H] = 0.44, [Ca/H] = $-0.15$ and [Sc/H] = $-0.59$.

\section{Coclusions}

Based on the catalogue of chemically peculiar stars of Ghazaryan et al. (2018) as well as on other published
works, we compiled a sample of 54 Am stars and obtained a homogeneous set of the Ca and Sc abundances with
taking into account departures from LTE.
The spectrum of HD~95608 retrieved from the archive UVES/VLT2 confirms the extremely low abundance
of Sc with the upper limit $[\textrm{Sc}/\textrm{H}]\le -1.64$ (NLTE).
At the same time the Ca and Fe abundances are typical for Am stars.

Our statistical analysis revealed correlations between the Ca, Sc abundances and the stellar effective temperature
$T_\mathrm{eff}$ with Spearman's rank correlation coefficients $\rho(\textrm{Ca}-T_\mathrm{eff})=0.54$
and $\rho(\textrm{Sc}-T_\mathrm{eff})=0.46$.
In stars with $\log g < 4$ the Ca and Sc abundances grow with increasing $T_\mathrm{eff}$ more rapidly than in stars
with $\log g \ge 4$.
Excess of Fe abundance with respect to the solar abundance on average is the same for
$7200\:\textrm{K} \le T_\mathrm{eff} \le 10300\:\textrm{K}$.
Deviations of the Ca and Sc abundances from their solar values do not correlate with [Fe/H] and the rotation
velocity.

Am stars exhibit on average the higher abundance [Ca/H] in comparison with [Sc/H] and
$[\textrm{Ca}/\textrm{Sc}]=0.41\pm 0.30$.
A scatter
of [Ca/Sc] in stars of similar
stellar parameters ($T_\mathrm{eff}$, [Fe/H], $V\sin i$) is large but,
for the hottest stars ($T_\mathrm{eff}>9500$\:K), there are some hints on systematic difference
between stars with $\log g \ge 4$ and $\log g < 4$.
Stars with $\log g \ge 4$ are characterized by a deficiency of Sc although it is smaller
than in cooler stars, and an enhancement of Ca relative to Sc remains:
$[\textrm{Ca}/\textrm{Sc}]\sim 0.5$.
In more evolved stars ($\log g < 4$) the chemical stratification is responsible for comparable excesses in
Ca, Sc and Fe superficial abundances.

Using the \textsc{MESA} we carried out calculations of stellar evolution with atomic diffusion for stars
with masses from 1.5 to $3M_\odot$.
Comparison with Ca and Fe abundances observed in Am stars of open clusters with the age greater than 600 Myr
(Praesepe, Coma, Hyades) allowed us to conclude that the theory agrees with observations and there is no need
for additional mechanisms responsible for separation or mixing of the stellar material.
Abundances of Ca and Sc in young Am stars of the Pleiades cluster with the age 110 Myr cannot be reproduced
simultaneously, independent of assumed stellar mass.

Elemental abundances of Am stars in four open clusters (Pleiades, Praesepe, Coma, Hyades) we compared with
turbulent models
by Richer et al. (2000) for Ca and Fe and by Hui-Bon-Hoa et al. (2022) for Sc as well as with mass loss models
by Vick et al. (2010) for Ca and Fe and by Hui-Bon-Hoa et al. (2022) for Sc.
Through these comparisons we arrived at the following conclusions.
The Fe abundances in Am stars with different values of the stellar mass and the age are reproduced by turbulent
models with the large turbulent diffusion coefficient $\omega=10^3$.
The same models predict almost constant Ca abundance for stars of the same age independently of the
stellar mass.
However observational data exhibit significant scatter (greater than 0.5 dex) for stars with close values of
$T_\mathrm{eff}$ and $\log g$ so that the models agree with observations of Ca only for a half of stars.
The scatter in observational data for Sc abundances is nearly the same as that for Ca abundances but we
arrived at the conclusion that the turbulent models with $\omega=500$ are more preferable than the models
with mass loss.
The diffusion models with mass loss  can be reconciled with Sc abundances observed in young stars only for
the mass loss rate $\dot M\sim 10^{-13}M_\odot/\textrm{yr}$.
For more evolved stars of the Praesepe and Hyades clusters the mass loss rate should be smaller by an order
of magnitude.

No one model with the mass and age of the well studied Am star Sirius reproduces its elemental abundances from
He to Ni.
The hypothesis that the chemical composition of Sirius was affected by mass transfer from the more evolved binary
companion seems to be improbable.

The mechanism of chemical peculiarity in atmospheres of Am stars arises from competition between gravitational
settling and radiative acceleration.
Nevertheless diffusion models especially of young stars need to be investigated in more detail with a more
thorough treatment of processes responsible for element separation in superficial layers.
The homogeneous data on calcium and scandium abundances obtained in the present study are useful for future tests
of more complicated models of chemically peculiar Am stars.

\section*{Funding}

This work was supported by ongoing institutional funding.
No additional grants to carry out or direct this particular research were obtained.

\section*{Conflict of interest}
The authors of this work declare that they have no conflicts of interest.

\section*{References}

\begin{enumerate}

\item H.A. Abt and K.I. Moyd, Astrophys. J. \textbf{182}, 809 (1973).

\item H.A. Abt and N.I. Morrell, Astrophys. J. Suppl. \textbf{99}, 135 (1995).

\item S.J. Adelman, Mon. Not. R. Astron. Soc. \textbf{266}, 97 (1994).

\item S.J. Adelman, Mon. Not. R. Astron. Soc. \textbf{280}, 130 (1996).

\item S.J. Adelman, Mon. Not. R. Astron. Soc. \textbf{310}, 146 (1999).

\item S.J. Adelman, H. Caliskan, D. Kocer, and C. Bolcal, Mon. Not. R. Astron. Soc. \textbf{288}, 470 (1997).

\item S.J. Adelman, H. Caliskan, T. Cay,  D. Kocer, and H.G. Tektunali, Mon. Not. R. Astron. Soc. \textbf{305}, 591 (1999).

\item S.J. Adelman,  A.F. Gulliver, and R. J. Heaton, Publ. Astron. Soc. Pacif. \textbf{127}, 58 (2015).

\item S.J. Adelman,  K.Yu, and A.F. Gulliver, Astron. Nachr. \textbf{332}, 153 (2011).

\item B. Baschek and A. Slettebak, Astron. Astrophys. \textbf{207}, 112 (1988).

\item E. B\"ohm-Vitense, Zeitschrift f{\"u}r Astrophys. \textbf{46}, 108 (1958).

\item J.M. Burgers, \textit{Flow equations for composite gases}, Academic Press, New York and London, 1969.

\item H. Caliskan and S.J. Adelman, Mon. Not. R. Astron. Soc. \textbf{288}, 501 (1997).

\item B. Campilho, M. Deal M., and D. Bossini, Astron. Astrophys. \textbf{659}, A162 (2022).

\item F. Castelli and S. Hubrig, Astron. Astrophys. \textbf{425}, 263 (2004).

\item G. Catanzaro, C. Colombo, C. Ferrara, and M. Giarrusso, Mon. Not. R. Astron. Soc. \textbf{515}, 4350  (2022).

\item P.S. Conti, Publ. Astron. Soc. Pacif. \textbf{82}, 781 (1970).

\item A.N. Cox, \textit{Allen's Astrophysical Quantities}, AIP Press, Springer (2000).

\item C.P. Folsom, O. Kochukhov, G.A. Wade, J. Silvester, and S. Bagnulo, Mon. Not. R. Astron. Soc. \textbf{407},  2383 (2010).

\item C.P. Folsom, G.A. Wade, and N.M. Johnson, Mon. Not. R. Astron. Soc. \textbf{433}, 3336 (2013).

\item L. Fossati, S. Bagnulo, R. Monier, S. A. Khan, O. Kochukhov, J. Landstreet, G. Wade, and W. Weiss,
      Astron. Astrophys. \textbf{476}, 911 (2007).

\item G.D. Gatewood and C.V. Gatewood, Astrophys. J. \textbf{225}, 191 (1978).

\item M. Gebran and R. Monier, Astron. Astrophys. \textbf{483}, 567 (2008).

\item M. Gebran, R. Monier, and O. Richard, Astron. Astrophys. \textbf{479}, 189 (2008).

\item M. Gebran, M. Vick, R. Monier, and L. Fossati, Astron. Astrophys. \textbf{523}, A71 (2010).

\item S. Ghazaryan, G. Alecian, and A.A. Hakobyan, Mon. Not. R. Astron. Soc. \textbf{480}, 2953 (2018).

\item F. Herwig, Astron. Astrophys. \textbf{360}, 952 (2000).

\item H. Hu, C.A. Tout, E. Glebbeek and M.-A. Dupret,  Mon. Not. R. Astron. Soc. \textbf{418}, 195 (2011).

\item A. Hui--Bon--Hoa, C. Burkhart, and G. Alecian, Astron. Astrophys. \textbf{323}, 901 (1997).

\item A. Hui--Bon--Hoa and G. Alecian, Astron. Astrophys. \textbf{332}, 224 (1998).

\item A. Hui--Bon--Hoa, G. Alecian, and F. LeBlanc, Astron. Astrophys. \textbf{668}, A6 (2022).

\item M. Jaschek and C. Jaschek, Astron. J. \textbf{62}, 343 (1957).

\item A.S. Jermyn, E.B. Bauer, J. Schwab, R. Farmer, W.H. Ball, E.P. Bellinger, A. Dotter, M. Joyce, P. Marchant,
      J.S.G. Mombarg, W.M. Wolf, W.T.L. Sunny, G.C. Cinquegrana, E. Farrell, R. Smolec, A. Thoul, M. Cantiello,
      F. Herwig, O. Toloza, L. Bildsten, R.H.D. Townsend, and F.X. Timmes, Astrophys. J. Suppl. Ser. \textbf{243}, 10 (2019).

\item V. Khalack and F. LeBlanc, Astron. J. \textbf{150}, 2 (2015).

\item T. K{\i}l{\i}{\c{c}}o{\u{g}}lu, R. Monier, J. Richer, L. Fossati, and B. Albayrak, Astron. J. \textbf{151}, 49 (2016).

\item O. Kochukhov, Astrophysics Source Code Library, record ascl:1805.015 (2018).

\item J. D. Landstreet, Astron. Astrophys. \textbf{528}, A132 (2011).

\item J. E. Lawler and J. T. Dakin, J. Opt. Soc. Am. B \textbf{6}, 1457 (1989).

\item J.E. Lawler, Hala, C. Sneden, G. Nave, M.P. Wood, and J.J. Cowan, Astrophys. J. Suppl. Ser. \textbf{241}, 21 (2019).

\item F. LeBlanc, V. Khalack, B. Yameogo, C. Thibeault, and I. Gallant, Mon. Not. R. Astron. Soc. \textbf{453}, 3766 (2015).

\item K. Lodders, Space Sci. Rev. \textbf{217}, id.44 (2021).

\item L. Mashonkina, Mon. Not. R. Astron. Soc. \textbf{527}, 8234 (2024).

\item L. Mashonkina, T. Ryabchikova, S. Alexeeva, T. Sitnova, and O. Zatsarinny, Mon. Not. R. Astron. Soc. \textbf{499}, 3706 (2020).

\item G. Michaud, Astrophys. J. \textbf{160}, 641 (1970).

\item G. Michaud, J. Richer, and M. Vick, Astron. Astrophys. \textbf{534}, A18 (2011).

\item P. Morel and Y. Lebreton, Astrophys. Space Sci. \textbf{316}, 61 (2008).

\item M. Netopil, L. Fossati, E. Paunzen, K. Zwintz, O. I. Pintado, and S. Bagnulo, Mon. Not. R. Astron. Soc. \textbf{442}, 3761 (2014).

\item M. Netopil, {\.I}. A. Oralhan, H. {\c{C}}akmak, R. Michel, and Y. Karata{\c{s}}, Mon. Not. R. Astron. Soc. \textbf{509}, 421 (2022).

\item E. Niemczura, S.J. Murphy, B. Smalley, K. Uytterhoeven, A. Pigulski, H. Lehmann, D.M. Bowman, G. Catanzaro,
      E. van Aarle, S. Bloemen, M. Briquet, P. De Cat, D. Drobek, L. Eyer, J.F.S. Gameiro, N. Gorlova, K. Kami{\'n}ski,
      P. Lampens, P. Marcos-Arenal, P. I. P{\'a}pics, B. Vandenbussche, H. Van Winckel, M. St{\c{e}}{\'s}licki, and M. Fagas,
      Mon. Not. R. Astron. Soc. \textbf{450}, 2764 (2015).

\item Yu.V. Pakhomov, T.A. Ryabchikova, and N.E. Piskunov, Astron. Rep. \textbf{63}, 1010 (2019).

\item B. Paxton, J. Schwab, E.B. Bauer, L. Bildsten, S. Blinnikov, P. Duffell, R. Farmer, J.A. Goldberg, P. Marchant,
      E. Sorokina, A. Thoul, R.H.D. Townsend, and F.X. Timmes, Astrophys. J. Suppl. \textbf{234}, 34 (2018).

\item G.W. Preston, Ann. Rev. Astron. Astrophys. \textbf{12}, 257 (1974).

\item N. Przybilla, M. Firnstein, M. F. Nieva, G. Meynet, and A. Maeder, Astron. Astrophys. \textbf{517}, A38 (2010).

\item P. Renson and J. Manfroid, Astron. Astrophys. \textbf{498}, 961 (2009).

\item J. Richer,  G. Michaud, and  S. Turcotte, Astrophys. J. \textbf{529}, 338 (2000).

\item A. Romanovskaya, T. Ryabchikova, Yu. Pakhomov, S. Korotin, and T. Sitnova, Mon. Not. R. Astron. Soc. \textbf{525}, 3386 (2023).

\item F. Royer,  J. Zorec, and A.E. G{\'o}mez, Astron. Astrophys. \textbf{463}, 671 (2007).

\item F. Royer, M. Gebran, R. Monier, S. Adelman, B. Smalley, O. Pintado, A. Reiners, G. Hill, and A. Gulliver,
      Astron. Astrophys. \textbf{562}, A84 (2014).

\item T. Ryabchikova, N. Piskunov, R.L. Kurucz, H.C. Stempels, U. Heiter, Y. Pakhomov, and P. S. Barklem, Phys. Scr. \textbf{90}, 054005 (2015).

\item M.J. Seaton, Mon. Not. R. Astron. Soc. \textbf{362}, L1 (2005).

\item T.M. Sitnova, L.I. Mashonkina, and T.A. Ryabchikova, Mon. Not. R. Astron. Soc. \textbf{477}, 3343 (2018).

\item C. Spearman, Am. J. Psychol. \textrm{15}, 72 (1904).

\item O. Trust, L. Mashonkina, E. Jurua, P. De Cat, V. Tsymbal, and S. Joshi, Mon. Not. R. Astron. Soc. \textbf{524}, 1044 (2023).

\item V. Tsymbal, T. Ryabchikova, and T. Sitnova,
      in Kudryavtsev D.O., Romanyuk I.I., Yakunin I.A., eds,
      Astron. Soc. Pacific Conf. Ser. \textbf{518}. Physics of Magnetic stars,
      San Francisco: Astronomical Society of the Pacific, 247 (2019).

\item S. Turcotte, J. Richer, G. Michaud, C.A. Iglesias, and F.J. Rogers, Astrophys. J. \textbf{504}, 539 (1998).

\item O. Varenne and R. Monier, Astron. Astrophys. \textbf{351}, 247 (1999).

\item K.A. Venn and D.L. Lambert, Astrophys. J. \textbf{363}, 234 (1990).

\item M. Vick, G. Michaud, J. Richer, and O. Richard, Astron. Astrophys. \textbf{521}, A62 (2010).

\item W.D. Watson, Astrophys. J. \textbf{162}, L45 (1970).

\end{enumerate}

\end{document}